\documentclass[namedreferences,hyperref,optionalrh]{spr-sola}
\usepackage{graphicx}        
\usepackage{color}           
\usepackage{ragged2e}
\usepackage{amssymb}
\usepackage{textgreek}
\usepackage{tablefootnote}
\usepackage{amsmath}
\usepackage{booktabs}
\usepackage[caption=false]{subfig}

\newcommand{\ion}[2]{#1\,\textsc{#2}}
\newcommand{\as}{$^{\prime\prime}$}

\newcommand{\degree}{\textsuperscript{o}}

\chardef\us=`\_

\newcommand{\aia}{AIA}
\newcommand{\asos}{ASO-S}
\newcommand{\dkist}{DKIST}
\newcommand{\eis}{EIS}
\newcommand{\eovsa}{EOVSA}
\newcommand{\eui}{EUI}
\newcommand{\fermi}{FERMI}
\newcommand{\goes}{GOES}

\newcommand{\hmi}{HMI}
\newcommand{\hri}{EUI/HRI}
\newcommand{\hrieuv}{EUI/HRI\textsubscript{EUV}}
\newcommand{\hrieuvcor}{\hrieuv~174\,\AA}

\newcommand{\hinode}{Hinode}
\newcommand{\hrt}{PHI/HRT}
\newcommand{\hxi}{HXI}
\newcommand{\iris}{IRIS}
\newcommand{\metis}{Metis}
\newcommand{\sdo}{SDO}
\newcommand{\solophi}{PHI}
\newcommand{\rhessi}{RHESSI}
\newcommand{\solo}{Solar Orbiter}
\newcommand{\solohi}{SoloHI}
\newcommand{\spice}{SPICE}
\newcommand{\stereo}{STEREO}
\newcommand{\stix}{STIX}

\newcommand{\trace}{TRACE}
\newcommand{\xrs}{GOES/XRS}
\newcommand{\xrt}{XRT}

\newcommand{\campaign}{\solo\ Major Flare campaign}

\newcommand{\gskfont}{
  \bfseries 
  \color{red}
}
\DeclareTextFontCommand{\gsk}{\gskfont}

\begin{document}

\begin{frontmatter}
\title{Solar Orbiter's 2024 Major Flare Campaigns: An Overview}


\author[addressref={affA,affB},email={daniel.ryan@ucl.ac.uk}]{\inits{D.F.}\fnm{Daniel F.}~\snm{Ryan}\orcid{0000-0001-8661-3825}}

\author[addressref={affC}]{\inits{L.A.}\fnm{Laura A.}~\snm{Hayes}\orcid{0000-0002-6835-2390}}
\author[addressref={affB,affD}]{\inits{H.}\fnm{Hannah}~\snm{Collier}\orcid{0000-0001-5592-8023}}
\author[addressref={affF,affG}]{\inits{G.S.}\fnm{Graham S.}~\snm{Kerr}\orcid{0000-0002-5632-2039}}

\author[addressref={affF,affG}]{\inits{A.R.}\fnm{Andrew R.}~\snm{Inglis}\orcid{0000-0003-0656-2437}}

\author[addressref=affL]{\inits{D.}\fnm{David}~\snm{Williams}\orcid{0000-0001-9922-8117}}
\author[addressref=affL]{\inits{A.P.}\fnm{Andrew P.}~\snm{Walsh}\orcid{0000-0002-1682-1212}}
\author[addressref={affM,affS}]{\inits{M.}\fnm{Miho}~\snm{Janvier}\orcid{0000-0002-6203-5239}}
\author[addressref=affM]{\inits{D.}\fnm{Daniel}~\snm{M\"uller}\orcid{0000-0001-9027-9954}}

\author[addressref={affH}]{\inits{D.}\fnm{David}~\snm{Berghmans}\orcid{0000-0003-4052-9462}}
\author[addressref={affH}]{\inits{C.}\fnm{Cis}~\snm{Verbeeck}\orcid{0000-0002-5022-4534}}
\author[addressref={affH}]{\inits{E.}\fnm{Emil}~\snm{Kraaikamp}\orcid{0000-0002-2265-1803}}

\author[addressref={affG,affJ}]{\inits{P.R.}\fnm{Peter R.}~\snm{Young}\orcid{0000-0001-9034-2925}}
\author[addressref={affG}]{\inits{A.}\fnm{Therese A.}~\snm{Kucera}\orcid{0000-0001-9632-447X}}

\author[addressref={affB,affE}]{\inits{S.}\fnm{Säm}~\snm{Krucker}\orcid{0000-0002-2002-9180}}
\author[addressref={affB,affD}]{\inits{M.Z.}\fnm{Muriel Z.}~\snm{Stiefel}\orcid{0000-0002-8538-3455}}

\author[addressref={affK}]{\inits{}\fnm{Daniele}~\snm{Calchetti}\orcid{0000-0003-2755-5295}}

\author[addressref={affN}]{\inits{K.K.}\fnm{Katharine K.}~\snm{Reeves}\orcid{0000-0002-6903-6832}}
\author[addressref={affP}]{\inits{D.}\fnm{Sabrina}~\snm{Savage}\orcid{0000-0002-6172-0517}}

\author[addressref={affQ,affR}]{\inits{}\fnm{Vanessa}~\snm{Polito}\orcid{0000-0002-4980-7126}}


\address[id=affA]{University College London, Mullard Space Science Labratory, Holmbury St Mary, Dorking, Surrey RH5 6NT UK}
\address[id=affB]{University of Applied Sciences and Arts Northwest Switzerland (FHNW), Bahnhofstrasse 6, Windisch 5210, Aargau, Switzerland}
\address[id=affC]{Astronomy \& Astrophysics Section, School of Cosmic Physics, Dublin Institute for Advanced Studies, DIAS Dunsink Observatory, Dublin, D15 XR2R, Ireland.}
\address[id=affD]{Institute for Particle Physics and Astrophysics (IPA), Swiss Federal Institute of Technology in Zurich (ETHZ), Wolfgang-Pauli-Strasse 27, 8039 Zurich, Switzerland}
\address[id=affF]{Department of Physics, Catholic University of America, 620 Michigan Ave NE, Washington, DC 20064, USA}
\address[id=affG]{NASA Goddard Space Flight Center, Heliophysics Science Division, 8800 Greenbelt Road, Greenbelt, MD 20771, USA}
\address[id=affL]{European Space Agency, European Space Astronomy Centre (ESAC), 28692 Villanueva de la Cañada, Madrid, Spain}
\address[id=affM]{European Space Agency, European Space Research and Technology Centre (ESTEC), Keplerlaan 1, 2201 AZ Noordwijk, The Netherlands}
\address[id=affS]{Université Paris-Saclay, CNRS, Institut d'Astrophysique Spatiale, 91405, Orsay, France}
\address[id=affH]{Royal Observatory of Belgium, Ringlaan -3- Avenue Circulaire, B1180 Brussels, Belgium}
\address[id=affJ]{Northumbria University, Newcastle upon Tyne, NE1 8ST, UK}
\address[id=affE]{Space Sciences Lab, UC Berkeley, 7 Gauss Way, Berkeley, CA 94708, USA}
\address[id=affK]{Max-Planck-Institut für Sonnensystemforschung, 37077 Göttingen, Germany}
\address[id=affN]{Center for Astrophysics, Harvard-Smithsonian, 60 Garden Street, Cambridge, MA, USA}
\address[id=affP]{NASA Headquarters Heliophysics Division, 300 E Street SW, Washington, DC 20546}
\address[id=affQ]{Lockheed Martin Solar \& Astrophysics Laboratory, Org. A021S, Bldg. 203, 3251 Hanover St., Palo Alto, CA 94304, USA}
\address[id=affR]{Department of Physics, Oregon State University, 301 Weniger Hall, Corvallis, OR 97331}

\runningauthor{Ryan et al.}
\runningtitle{Solar Orbiter's Major Flare Campaigns of 2024: An Overview}

\begin{abstract}

\solo\ conducted a series of flare-optimised observing campaigns in 2024 utilising the Major Flare Solar Orbiter Observing Plan (SOOP).
Dedicated observations were performed during two distinct perihelia intervals in March/April and October, during which over 22 flares were observed, ranging from B- to M-class. These campaigns leveraged high-resolution and high-cadence observations from the mission’s remote-sensing suite, including the High-Resolution EUV Imager (\hrieuv), the Spectrometer/Telescope for Imaging X-rays (\stix), the Spectral Imaging of the Coronal Environment (SPICE) spectrometer, and the High Resolution Telescope of the Polarimetric and Helioseismic Imager (\hrt), as well as coordinated ground-based and Earth-orbiting observations.
\hrieuv\, operating in short-exposure modes, provided two-second-cadence, non-saturated EUV images, revealing structures and dynamics on scales not previously observed.
Simultaneously, \stix\ captured hard X-ray imaging and spectroscopy of accelerated electrons, while SPICE acquired EUV slit spectroscopy to probe chromospheric and coronal responses.
Together, these observations offer an unprecedented view of magnetic reconnection, energy release, particle acceleration, and plasma heating across a broad range of temperatures and spatial scales.
These campaigns have generated a rich dataset that will be the subject of numerous future studies addressing Solar Orbiter’s top-level science goal: {\it ``How do solar eruptions produce energetic particle radiation that fills the heliosphere?''}.
This paper presents the scientific motivations, operational planning, and observational strategies behind the 2024 flare campaigns, along with initial insights into the observed flares.
We also discuss lessons learned for optimizing future \campaign s and provide a resource for researchers aiming to utilize these unique observations.

\end{abstract}
\keywords{Integrated Sun Observations; Flares; Active Regions; Corona, Active; X-Ray Bursts}
\end{frontmatter}

\section{Introduction}
Solar flares are the most energetic events in our solar system, resulting in the rapid release of magnetic energy in the solar atmosphere  \citep{Fletcher:2011, benz_2017}.
Within seconds to minutes, this energy is converted into plasma heating, particle acceleration, bulk plasma flows, and the generation of radiation across the electromagnetic spectrum.
Solar flares are a fundamental component of solar eruptive events, often occurring in conjunction with coronal mass ejections (CMEs) and solar energetic particles (SEPs), which can drive adverse space weather conditions at Earth and throughout the solar system.
Despite significant observational and theoretical advances in recent years, fundamental questions remain concerning the underlying physics driving the magnetic reconnection, efficient particle acceleration, energy dissipation, the spatial and temporal evolution of flare plasma at the finest scales, and the three-dimensional evolution of flare structures within the broader magnetic topology.



Addressing these questions requires new approaches that leverage multi-scale and multi-perspective observations of solar flares.
The Solar Orbiter mission \citep{solo}, launched in 2020, offers such an approach by combining remote-sensing and in-situ instruments with recurrent vantage points away from the Sun-Earth line and frequent close proximity to the Sun.
This provides an unprecedented opportunity to study solar flares from a new perspective. 
However, the mission’s limited telemetry imposes constraints on the volume of data that can be transmitted to Earth.
To overcome this, Solar Orbiter employs Solar Orbiter Observing Plans (SOOPs).
These are targeted campaigns that prioritise specific scientific objectives by coordinating instruments' observing programs and concentrating spacecraft resources \citep{Zouganelis_2020}.
This ensures that the highest-value data are collected and downlinked.
SOOPs also enable coordination with Earth-based and near-Earth observatories, leveraging their complementary capabilities and viewing locations.
By focusing resources on specific scientific objectives and tailoring observational strategies, SOOPs maximize the scientific return of the mission within the telemetry constraints.

The Major Flare SOOP\footnote{\textit{L\_BOTH\_HRES\_HCAD\_Major\-Flare}: \url{https://s2e2.cosmos.esa.int/confluence/display/SOSP/L_BOTH_HRES_HCAD_Major-Flare}} is designed to capture high-resolution, high-cadence observations of solar flares with the aim of investigating the spatial and temporal evolution of flare emission.
It combines flare-optimised observations of multiple remote-sensing instruments, including high-cadence, high-resolution, short exposure images in the extreme ultraviolet (EUV) by the Extreme Ultraviolet Imager's High Resolution Imager \citep[\hrieuv;][]{eui}, high-resolution EUV spectral diagnostics from the Spectral Imaging of the Coronal Environment spectrometer \citep[\spice;][]{spice}, and hard X-ray (HXR) observations from the Spectrometer/Telescope for Imaging X-rays \citep[\stix;][]{stix}.
Optical images and magnetograms of the photosphere are provided by the Polarimetric and Helioseismic Imager \citep[\solophi;][]{phi}.
Finally, all in-situ instruments operate during the Major Flare SOOP, and can help characterise flare-related phenomena that pass the spacecraft, e.g. SEPs.

The March/April 2024 \campaign\ was the first to successfully observe flares with the Major Flare SOOP.
It was followed by a repeat campaign in October 2024.
Across both campaigns, at least 22 flares from B- to M-class were observed by a subset of \solo's instruments.
Several were also jointly captured by Earth-based and near-Earth observatories thanks to coordinated observing campaigns.
The resulting unique datasets provide a comprehensive view of flare evolution, encompassing signatures of magnetic reconnection, energy release, and particle acceleration, and can help to address some key outstanding questions in solar flare physics.

\solo\ has observed flares outside of the above-mentioned campaigns.
\stix\ has observed tens of thousands of flares thanks to its near 100\% duty cycle, which includes many times when no specific SOOP is being run.
Other \solo\ campaigns have captured flare observations too.
Examples include the M7 flare of 2024-09-30 observed during the L\_SMALL\_MRES\_MCAD\_Earth-Quadrature\footnote{\url{https://s2e2.cosmos.esa.int/confluence/display/SOSP/L\_SMALL\_MRES\_MCAD_Earth-Quadrature}} SOOP, and the multiple flares observed during the December 2024 flare campaign.
These campaigns have provided valuable flare observations that will help improve our understanding of solar flares.
That said, their specific science goals were subtlely different and resulted in different combinations of instrument observing modes.

In the interest of scope, therefore, this paper focuses on the March/April and October 2024 \campaign s.
We outline the campaigns' scientific goals (Section~\ref{sec:science}), operational planning (Section~\ref{sec:planning}), observations (Section~\ref{sec:obs-summary}), initial findings (Section~\ref{sec:results}), and a brief discussion of lessons learned that may be useful for future campaigns (Section~\ref{sec:disc}). 
Given the volume of high-quality observations, detailed analysis of specific events is left to dedicated future papers. 
Instead, this paper provides an overview of the campaigns that serves as context for those detailed studies, promotes further use of these observations within the solar physics community, and highlights some of \solo\ contributions to our study of solar flares.

\section{Science Goals of \solo\ Major Flare Campaign}
\label{sec:science}
The three primary science goals of the 2024 \campaign s were to:

\begin{enumerate}
\item \textbf{Probe the spatial evolution of flare plasma on timescales relevant to impulsive energy release.} This was achieved by imaging flares in EUV at 174~\AA, representing the highest temporal (2\,s) and spatial (200\,--\,300\,km) resolution observations of the flaring EUV corona to that date.
This was complemented by \stix\ HXR spectroscopic imaging with cadences as low as 0.5 s. High-cadence (5\,s) EUV `sit-and-stare' spectroscopic observations from SPICE added further insights into the plasma response along the slit location.
\item \textbf{Detect the presence of flare-accelerated protons via the Orrall-Zirker effect}. \spice\ observations were designed to search for evidence of charge-exchange between flare accelerated protons and ambient  H~\textsc{i}. This would reveal itself as a broad, redshifted emission of the H~\textsc{i} Ly$\beta$ spectral line and nearby continuum.
\item \textbf{Explore the three-dimensional geometry of flare sources and their evolution over time.} This goal takes advantage of Solar Orbiter’s unique vantage point, with simultaneous observations from Solar Orbiter and ground-based/Earth-orbiting telescopes at a Solar Orbiter–Sun–Earth separation angle of 90$^{\circ}$ $\pm$ 20$^{\circ}$.
\end{enumerate}

These science goals are not exhaustive, and other goals can be address by the campaigns' observations.
However, these were the most influential in driving the campaigns' design and operational decisions.
We shall therefore briefly discuss them in the subsections below.

\subsection{Probing the Spatial Evolution of Flare Plasma on Timescales Relevant to Impulsive Energy Release}

The acceleration of non-thermal electrons and plasma heating are among the earliest observable consequences of impulsive magnetic energy release in the solar corona. HXR ($\gtrsim$\,4\,keV) observations provide the most direct diagnostic of flare-accelerated electrons and the hottest flare plasma ($>$10~MK). The non-thermal bremsstrahlung HXR spectrum directly reflects the energy distribution of accelerated electrons, and hence it can be inverted to reveal insights into the acceleration and transport process(es) that produced it. HXR imaging therefore reveals the sites of energy release and/or dissipation.
That said, HXR bremsstrahlung observations have limitations.
Bremsstrahlung produces insufficient HXRs in non-flaring conditions to reveal the background topology in which flares occur, and the typical angular resolution achieved by previous and current HXR spectroscopic imagers (a few arcseconds) is insufficient to resolve individual magnetic structures.
These facets are currently better observed via high-resolution EUV imaging.
Consequently, HXR spectroscopic imaging and high resolution EUV imaging are a powerful combination for elucidating the processes that drive solar flares.

Modelling in recent decades suggests that non-thermal electron acceleration in solar flares occurs on timescales of a few seconds or less \citep[e.g.][]{drake:2006,drake:2013}.
This is consistent with spatially integrated observations of variations down to similar timescales in HXR signatures of accelerated electrons \citep[e.g.][]{Collier:2023,Collier:2024a,Inglis:2024}.
Despite this, neither HXR spectroscopic imaging nor unsaturated, high-resolution EUV imaging have been previously available on such timescales.
These limitations can be overcome in the HXR-regime by \stix, which can image bright flares as fast 0.3\,--\,0.5\,s, a factor of 4\,--\,10 faster than its predecessor, \rhessi\ \citep[][]{rhessi}.
In the EUV imaging regime, \hrieuv\ has a configurable exposure time which can achieve imaging cadences of $\leq$2\,s, a factor of 6 faster than the nominal \sdo/\aia\ cadence of 12s.
The short exposure mode of \hrieuv\ (Section~\ref{sec:eui_mode}) means that we can identify flaring sources that typically saturate detectors.
Moreover, its CMOS Active Pixel Sensor (APS) detectors do not suffer from the same blooming effects during saturation as the CCD detectors used by \sdo/\aia. 
Finally, \hrieuv\ has a two-pixel angular resolution of 1'', an improvement over \sdo/\aia's 1.5''.
This translates to a comparable spatial resolution at perihelion (200\,km) to that of the Hi-C sounding rocket \citep[0.3\,--\,0.4'' $\equiv$ 220\,--\,290\,km;][]{hic1}.
This makes \hrieuv\ the most suitable currently available satellite-based EUV imager that can be configured for solar flares observations.
By combining these observations with spectroscopic observations from \spice, we can leverage multiple plasma diagnostics (temperature, mass flows, microturbulence and composition; albeit, at a lower spatial resolution and from a single slit).
These can probe the plasma response to flare energy injection on timescales of $\sim$\,5\,s across a wide temperature range ($\log T \sim 4 - 7$; chromosphere through corona).

By combining these capabilities, \solo's Major Flare SOOP represents an unprecedented advance in our ability to address open science questions regarding the spatial evolution of plasma structures and accelerated electrons in solar flares on timescales approaching those of the underlying energy release processes.


\subsection{The Orrall-Zirker Effect}
\label{sec:oz-goal}
The majority of our knowledge on flare-accelerated charged particles comes from measurements of accelerated electrons through HXR observations.
By contrast, the role and energy budget of accelerated ions remains relatively unexplored, despite the fact that they may share a common acceleration mechanism with electrons \citep[e.g.][]{Shih:2009}, and could transport as much, or even more, energy than accelerated non-thermal electrons \citep[][]{Yin:2024}.
Omitting them from our models and theories therefore represents an important gap in our ability to understand flare energetic processes. 
 
Part of why accelerated ions are often ignored is the difficulty in characterising their non-thermal distribution.
Diagnosing protons with energies $>$\,1\,MeV requires $\gamma$-ray observations, of which only a limited number have been obtained by instruments such as \rhessi\ and \fermi\ \citep[e.g.\ ][]{Shih:2009,Pesce-Rollins:2024}.
An alternative for diagnosing proton energies $<$\,1\,MeV, where the bulk of the energy content may reside, is the so-called Orrall-Zirker effect \citep[][]{orrall:1976}.
Accelerated deka- and hecta-keV protons within flaring loops can undergo charge-exchange interactions with the ambient neutral hydrogen in the chromosphere.
This creates energetic neutral hydrogen atoms, some of which end up in an excited state and ultimately emit a photon.
When observed near disk-centre, downward accelerated protons and their associated energetic neutral H atoms travel away from the observer.
Their emission should therefore appear as a broad redshifted feature in the wings of H\,\textsc{i} Lyman lines\footnote{The equivalent feature in the Balmer lines is likely too weak owing to the strength of the optical continuum.}.
The shape of this feature reveals information about the non-thermal proton distribution \citep[e.g.][]{Brosius:1999}.
Recently, \cite{Kerr:2023} revisited this process using modern numerical models that track both the non-equilibrium ionisation fraction during the flare, and the evolving non-thermal proton distribution.
Updated atomic cross-sections were used to determine the viability of this emission as a diagnostic of non-thermal protons in flare footpoints.
They found that while significantly more transient than earlier predictions assumed (due to rapid ionization of the chromosphere), the redshifted emission feature should be present for $\sim$\,1\,--\,5\,s.
With a sufficiently short exposure time, \spice\ should be capable of detecting this feature in the Ly$\beta$ line.

\subsection{Explore the 3D geometry of flare sources and their temporal evolution}
\label{sec:3d-goal}
The morphology and dynamics of high-energy sources are integral components of the evolution and energetics of solar flares.
Their locations can reveal where the hottest plasma resides within the 3D magnetic field structure, and their motions indicate something about the continued evolution of that structure.
Prior to \solo, the 3D properties of solar X-ray sources could not be determined due to the lack of co-temporal solar X-ray imaging observations at angular separations sufficient for 3D reconstruction.
This restricted height and velocity measurements to the plane of sky, which were only representative when the sources were observed near the solar limb and moved radially away from the Sun.
The absence of volume measurements has hindered the calculation of densities at higher temperatures and over broader temperature ranges than are accessible via density-sensitive emission line ratios.
Moreover, these are limited by assumptions (e.g.\ statistical equilibrium) that are typically invalid in flaring conditions.
This has made it challenging to reliably analyse physical processes and properties influenced by density, such as cooling processes \cite[e.g.][]{Cargill:1995,Ryan:2013} and flare energetics \citep[e.g.][]{Emslie:2012,Aschwanden:2015}.

However, since the launch of \solo, it has been shown \citep[][]{ryan:2024a,Ryan:2024b} that the 3D geometry of flare X-ray sources can be determined and charted through time by combining observations from \stix\ and \hinode's X-ray Telescope \citep[\xrt;][]{xrt} or the Hard X-ray Imager \citep[\hxi;][]{hxi} onboard the Advanced Space-based Solar Observatory \citep[\asos;][]{asos,asos2}.
3-D analyses of other flares, also involving EUV observations, are required before we can fully understand the complexity of high energy flare processes in 3D.

\begin{figure}
    \centering
    \includegraphics[width=0.5\textwidth]{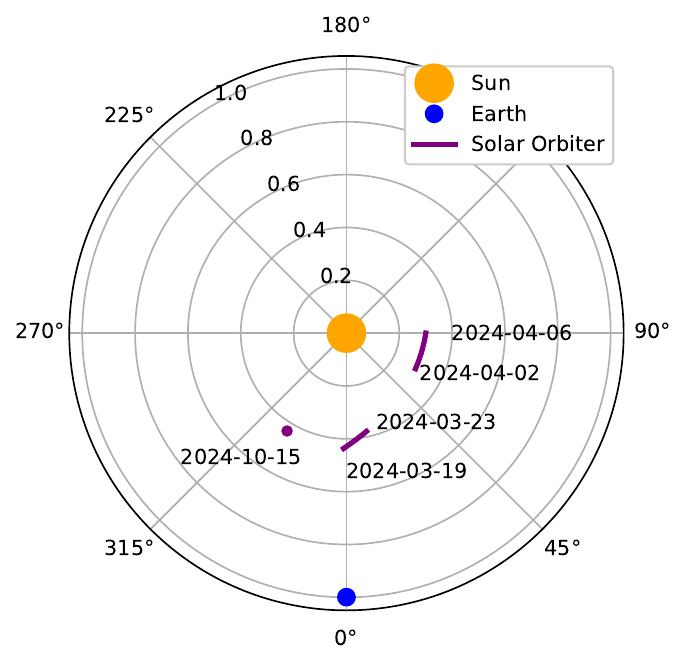}
    \caption{Heliographic Stonyhurst positions of Solar Orbiter on days of the various \campaign\ observing windows.}
    \label{fig:solo_locations}
\end{figure}

\begin{table}[b!]
    \caption{Summary of the observing windows of the \campaign.  Observing angle and distance represent values in the middle of the window.}
    \label{tab:windows}
    \centering
    \begin{tabular}{r|c|c|c|c}
         Date  & Interval [UT]  & \shortstack{Observing\\Angle\textsuperscript{a}}   & Distance\textsuperscript{b} [au] &\shortstack{Light Travel Time\\Difference\textsuperscript{c} [min:sec]} \\
         \hline
         2024 March 19    & 20:00\,--\,00:00  & 1.29\degree & 0.435 & 4:40    \\
         2024 March 23/24 & 22:30\,--\,02:30  & 3.33\degree & 0.381  & 5:08  \\
         2024 April 2     & 20:00\,--\,00:00  & 63.05\degree & 0.294  & 5:52 \\
         2024 April 4     & 20:00\,--\,00:00  & 76.45\degree & 0.293  & 5:53 \\
         2024 April 5     & 20:00\,--\,00:00  & 83.08\degree & 0.296  & 5:52 \\
         2024 April 6     & 20:00\,--\,00:00  & 89.54\degree & 0.300  & 5:50 \\

         \midrule
         2024 Oct 15     & 15:00\,--\,00:00\textsuperscript{*}  & -31.77\degree & 0.430  & 4:41 \\
    \end{tabular}
    \begin{FlushLeft}
        \footnotesize{\textsuperscript{a}Defined as the angle Earth center---Sun center---\solo\ position.\\
        \textsuperscript{b}Defined as distance between Sun centre and \solo\ in the middle of the observing window.\\
        \textsuperscript{c}Time difference between light emitted from Sun-centre reaching \solo\ and Earth. Calculated for the centre of the observing window and rounded to the nearest second.\\
        \textsuperscript{*} \hrieuv\ ran only a subset of this time window between 18:00-22:00 UT
        }
    \end{FlushLeft}
\end{table}

\section{Operational Planning}
\label{sec:planning}

\subsection{Campaign Scheduling}
The 2024 \campaign s took place March and April 2024 during remote sensing windows (RSWs) 13 and 14\footnote{\url{https://s2e2.cosmos.esa.int/confluence/display/SOSP/Solar+Orbiter+Planning+-+RSWs+13\%2C+14\%2C+15}}, and on 15th October during RSW 17.
These included the approach to, and transition through, the 5th and 6th perihelia of \solo's nominal mission phase.
Solar activity was expected to be substantially higher during these RSWs than previous oness due to the evolution of the solar cycle.

The campaigns were run over seven 4-hour observing windows (Table~\ref{tab:windows}).
Their duration was limited by the period for which \hrieuv\ can run high resolution, high cadence observations before filling up its data buffer.
After this, several hours are required for \eui\ to process the data and pass it to the spacecraft's solid-state memory, before high-cadence, high-resolution observations can once again be taken.
(Note that during the October 15th campaign \spice\ observed for an additional 4 hours outside the \hrieuv\ window.)
Additional observing windows could not be accommodated due to the scheduling requirements of other SOOPs.
The first two observing windows were held in March to approximately coincide with Solar Orbiter crossing the Sun-Earth line.
This maximised the chance of coordinated observations from Earth.
Four more were held in the first week of April, when Solar Orbiter's observing angle relative to Earth was in the range 60\degree\,--\,90\degree, and hence suitable for stereoscopy studies.
Finally, one window was held in October, when \solo\ was $\sim$-30\degree\ from the Sun-Earth line (see Figure \ref{fig:solo_locations}).

\subsection{Solar Orbiter Observing Modes}
\label{sec:obs-modes}
In order to achieve the science goals listed in Section~\ref{sec:science}, the campaign involved explicit coordination between four of \solo's remote sensing instruments: \hrieuv, \spice, \stix\ and the High Resolution Telescope of the Polarimetric and Helioseismic Imager \citep[\hrt;][]{phi,phihrt}.
\metis\ and \solohi\ were not utilised in this campaign because of the probable need to point \solo\ away from Sun-centre to place the target active region in the limited FOVs of \hrieuv, \spice\ and \hrt.
In-situ instruments were operated nominally during the campaign.

\subsubsection{\hrieuv}
\label{sec:eui_mode}
\hri\ provides high-resolution, high-cadence images in the 174\,\AA\ and Lyman-$\alpha$ passbands.
To maximise temporal cadence and bypass performance issues with the Lyman-$\alpha$ telescope near perihelion, only the 174\,\AA\ channel (\hrieuv) was used in these campaigns.

A crucial requirement for EUV observations of large flares is to avoid saturation.
Large flares can produce high fluxes from tiny localized regions, which can cause the number of photo-electrons in the photodiodes of corresponding pixels to approach the so-called {\it full well}.
In this regime, the measured digital response becomes increasingly non-linear and eventually saturates at a fixed value.
Further increasing the EUV flux causes the photodiodes of CCD detectors to overflow into neighbouring pixels, an effect is known as blooming, which has long hampered the observation of large flares by EUV imagers, like \aia.
One strength of \hrieuv\ for flare observations is its use of CMOS detectors, which are far less susceptible to blooming.
(That said, intense localised flare signal can still contaminate neighbouring CMOS pixels through diffraction on the telescope front filter grid or through the point-spread-function of the optics.)

\begin{figure}
    \centering
    \includegraphics[width=0.95\textwidth]{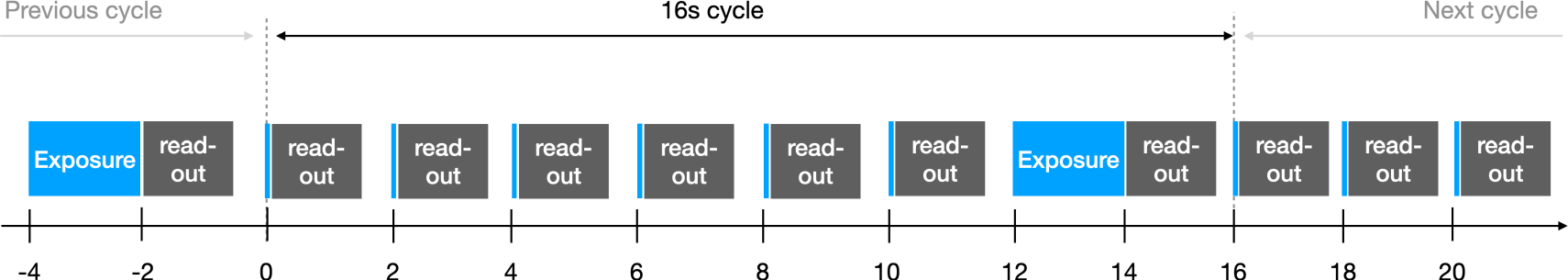}
    \caption{Schematic of the \hrieuv\ observing mode used in the 2024 \campaign s. Six short exposure (0.04\,s) images are followed by one long exposure (2\,s), before the cycle repeats. Accounting for detector readout time, this results in a 16\,s long exposure cadence and a 2\,s short exposure cadence (except when the long exposure image is taken).}
    \label{fig:eui_cadence}
\end{figure}

Another important consideration for EUV flare observations is the instrument's dynamic range, i.e.\ the difference between the brightest and faintest signals that can be displayed.
For an EUV camera, this is defined as the ratio of the full well capacity (or, more restrictively, the level at which the pixel response becomes non-linear) and the read-out noise.
To handle flare intensities beyond their dynamic range, previous solar EUV imagers including \trace\ and \aia\ used an automatic exposure-control algorithm that decreased the exposure time for every second image in response to a detected flare.
Thus, they increased dynamic range at the expense of cadence.

In order to minimise saturation and maximise cadence and dynamic range, this campaign leveraged \hrieuv's CMOS detectors and its programmable exposure time.
A cycle of 6 short (0.04\,s) and 1 long (2\,s) exposure images was predefined (Figure~\ref{fig:eui_cadence}).
This enabled a long exposure cadence of 16\,s and a short exposure cadence of 2\,s (except when the long exposure images were being taken).
This combination allows the very bright flare kernels to be imaged at 2s cadence with the short exposures, while imaging fainter features with the long exposures images.

The short exposure images are strongly compressed (to a few percent of the regular images), allowing the 2s-cadence can be sustained for 4 hours, even with \solo's telemetry-constraints \citep{Collier:2024b}.
During the March 19 and 23/24 windows, a lossless compression scheme was used, and only pixels with DN values near or above the maximum of the long exposure images were telemetered.
During the April windows, a lossy compression scheme was used, in which all pixel values were telemetered.
This allowed more information about the dimmer plasma to be retained.
(See Section~\ref{sec:disc} for a comparison of observations obtained with the two compression schemes.)

\subsubsection{\spice}
\label{sec:spice_mode}
\spice\ was used to probe the rapid evolution of the plasma at multiple temperatures and altitudes, and to search for signs of non-thermal protons.
Table~\ref{tab:spectraoverview} provides of full list of the spectral windows observed by \spice\ during the campaign.
Among them are windows around the blue wing of Ly$\beta$, the full Ly$\beta$ line itself, and the red wing of the line including the region around O\,\textsc{vi} 103.2\,nm and O\,\textsc{vi} 103.7\,nm.
This was to search for signs of excess red wing emission compared to the blue wing $\lambda\sim[8-15]$~\AA\ from Ly$\beta$ line core, as predicted by the Orrall-Zirker effect.
Other lines from the \spice\ short wavelength channel were chosen to sample the transition region and hot flare lines.
To maximise cadence, a sit-and-stare mode with the 4$^{\prime\prime}$ slit was used with a $4.8$~s exposure time, and at an overall cadence of $5.1$~s.
This represents the upper edge of detectability of the Orrall-Zirker effect according to the simulations of \citet{Kerr:2023}.
A context raster was made prior to the start of sit-and-stare observations.

\begin{table*}
\caption{\spice\ observing windows for the 2024 \campaign s.} 
\label{tab:spectraoverview}     
\centering                                     
\begin{tabular}{l l r c }          
\hline                       
Window ID & Ion & $\lambda$~[nm] & $\log T$~[K] \\    
\hline                                  
Ly-g-CIII group (Merged)       & H~\textsc{i} Ly$\gamma$            &  $97.254$   & $\sim$4.0 \\ 
Ly Beta 1025 (Merged)          & H~\textsc{i} Ly$\beta$             &  $102.572$  & $\sim$4.0 \\ 
O VI 1037 (Merged)             & C~\textsc{ii}                      &  $103.634$  & 4.39  \\ 
Fe XX 721 + O II (Merged)      & O~\textsc{ii}                      &  $71.850$   & 4.45 \\ 
Ly beta blue continuum (Merged)& S~\textsc{iii}\textsuperscript{a}  &  $101.249$  & 4.70 \\ 
Ly beta blue continuum (Merged)& S~\textsc{iii}\textsuperscript{a,b}&  $101.550$  & 4.70 \\ 
Ly beta blue continuum (Merged)& S~\textsc{iii}\textsuperscript{a,b}&  $101.578$  & 4.70 \\
Ly-g-CIII group (Merged)       & C~\textsc{iii}                     &  $97.702$   & 4.85 \\
O III 703 / Mg IX 706 (Merged) & O~\textsc{iii}\textsuperscript{b}  &  $70.385$   & 4.90 \\  
N IV 765 - Peak                & N~\textsc{iv}                      &  $76.515$   & 5.15 \\  
O VI 1032 (Merged)             & O~\textsc{vi}                      &  $103.191$  & 5.45  \\ 
O VI 1037 (Merged)             & O~\textsc{vi}                      &  $103.761$  & 5.45  \\ 
Ne VIII 770 - Peak             & Ne~\textsc{viii}                   &  $77.044$   & 5.80  \\ 
O III 703 / Mg IX 706 (Merged) & Mg~\textsc{ix}                     &  $70.606$   & 6.00 \\  
Ly-g-CIII group (Merged)       & Fe~\textsc{xviii}                  &  $97.486$   & 6.85  \\ 
Fe XX 721 + O II (Merged)      & Fe~\textsc{xx}                     &  $72.156$   & 7.00 \\            

\hline            
\end{tabular}
\begin{FlushLeft}
\footnotesize{
The Window ID refers to the label within the \spice\ files. Wavelengths and ionisation equilibrium formation temperatures are taken from the CHIANTI atomic database \citep{DelZanna:2015}, via FIASCO \citep{Barnes2024}.\\
\textsuperscript{a}{These lines only appeared above the background in flare footpoints (i.e. during transient heating).}\\
\textsuperscript{b}{These lines are strongly blended.}}
\end{FlushLeft}
\end{table*}

To save telemetry, on-board spectral summing of (2x1) was performed for every window, except the `Ly beta blue continuum (Merged)' window that used (4x1) summing.
This gives spectral plate scales of $\delta\lambda = 0.0195034 $~nm for the short wavelength (SW) windows ($70.4-79.0$~nm), $\delta\lambda = 0.019246 $~nm for the long wavelength (LW) windows ($97.3-104.9$~nm), with the exception of `Ly beta blue continuum (Merged)' that had $\delta\lambda = 0.038492$~nm.
It has been estimated that the spectral resolution for the SW channel is 7.8 pixels, and for the LW channel 9.4 pixels \citep{Fludra:2021}.
Along the slit, the spatial plate scale is $1.09798^{\prime\prime}$, with the FWHM of the point spread function estimated to be 6.3 pixels \citep{Fludra:2021}.
Additionally, \spice\ has a complicated point spread function (PSF) that is tilted in the spatial-spectral plane, with the result being that Doppler shift artefacts can appear where there are strong intensity gradients \citep{Fludra:2021,Plowman:2023}.
Work is ongoing to characterise this PSF \citep{Plowman:2023}, but as noted in \spice's Data Release 5 \citep{spicedr5}, analysis of Doppler signals is currently possible, but non-trivial.
Strong Doppler shifts ($>$200\,km/s) associated with erupting structures, were characterised in \cite{janvier:2023} in the context of an M-class flare, and co-observed with \hinode/\eis.
Advice should be sought from the \spice\ instrument team when interpreting Doppler shifts or line widths.
Level-2 (science ready) data is now available via the SOAR\footnote{Solar Orbiter Archive: \url{https://soar.esac.esa.int/soar/}}, as part of \spice\ DR5. 

\subsubsection{\stix\ \& \hrt}
\label{sec:stix_phi_mode}
\stix\ is unique among \solo's remote sensing instruments in that it is (almost) always observing, even outside RSWs.
It enables spatial and spectral characterisation of electron acceleration and plasma heating on cadences that can be as fast as 0.5\,s, depending on the incident photon flux.
Thanks to its large FOV and low telemetry requirement, \stix's nominal operating mode was sufficient for the campaign.
It thus provided reliable continuous observations with minimal coordination complexity.

\hrt\ provided high-resolution, low-cadence (30\,minute) photospheric spectro-polarimetric context data. 
These enabled the full Stokes vector to be inverted with the Milne-Eddington RTE inversion code MILOS \citep{milos}.
From this, the full magnetic field vector, line-of-sight velocity and continuum intensity could be retrieved \citep[more information on the data reduction can be found in][]{hrtpipeline}. 
The overall quality of the data depends on the location of the observed active region.
In particular, the polarimetric noise is higher close to the limb and the optical aberrations are slightly stronger.

\begin{figure}
    \centering
    \includegraphics[width=0.95\textwidth]{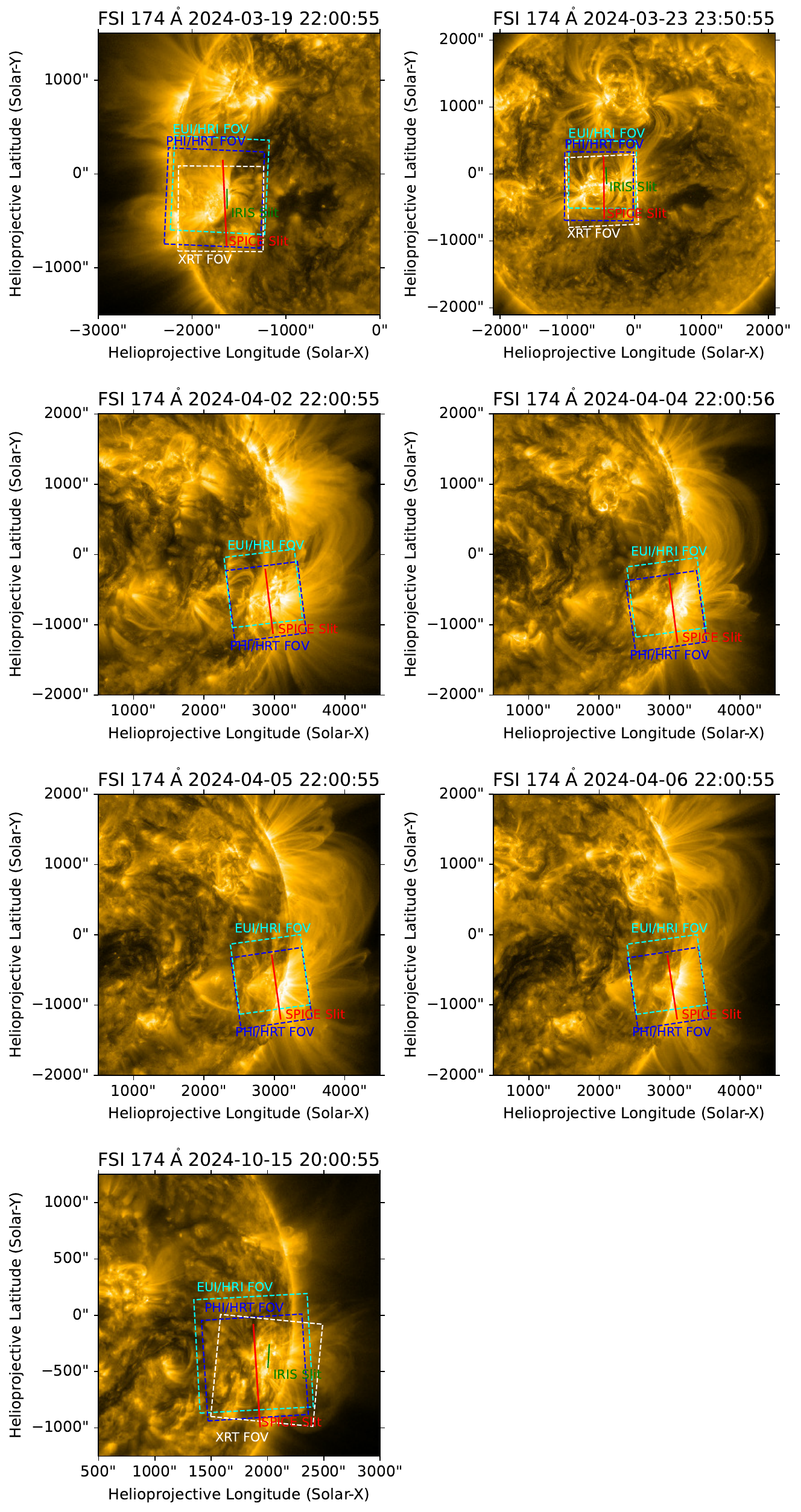}
    \caption{Fields of view of instruments used in the \campaign\ for each observing window.  They are overlaid on the \eui/FSI~174\,\AA\ image closest to the middle of the window.  Full-disk instruments are not shown. }
    \label{fig:fovs}
\end{figure}

\subsection{Targeting}
\label{sec:targeting}
The 2024 \campaign s had to be planned long before it was known whether there would be a flaring active region visible.
Fortunately, the Sun supplied such active regions during the campaigns.
The March/April campaign targeted NOAA active region (AR) 13615, while the October campaign targeted NOAA AR~13852.
Both were highly flare-productive.
For SOOPs that require a specific pointing with a time-evolving target, pointing decisions can be made up to 48 hours before the observation time, depending on the spacecraft pass schedule.
This was the case for the 2024 \campaign s, which enabled the pointing to be adjusted based on the AR location and predicted passage across the solar disk.

Figure~\ref{fig:fovs} shows the fields of view (FOVs) of instruments used during each observing window (fine-scale pointing adjustments, e.g.\ due to instrumental effects, are not accounted for).
They are overlaid on the \eui/Full Sun Imager (FSI)~174\,\AA\ image closest in time to the middle of the observing window.
Instruments that have full-Sun FOVs are not shown.
NOAA AR 13615 was visible in front of the east limb and near disk centre on March 19 and 23/24, respectively, from both \solo\ and Earth.
By the April observing windows, the active region had rotated beyond the west limb as seen from Earth.
However, it remained visible to \solo\ because its orbit caused it to quasi-co-rotate with the Sun during that time.
It only passed beyond the west limb as seen by \solo\ shortly after the last observing window.
While a target active region that was visible to \solo\ and Earth would have been preferable, there was no such active region.
Hence, AR 13615 remained the target for the whole flare campaign, and successful coordinated observations with Earth were only made during the March windows.

For the 2024 October 15 instance of the Major Flare SOOP, AR 13852 was chosen as the target. AR 13854 was also in close proximity and was included in the EUI FOV. At the time of observation, Solar Orbiter was trailing the Earth view by approximately $32^\circ$, thus the target active region was clearly on-disk for Earth-based instrumentation.

\subsection{Coordination with Other Observatories}
\label{sec:coord-obs}

\subsubsection{\hinode: \xrt\ \& \eis}
\label{sec:coord-hinode}
Coordination with \hinode\ involved two of its instruments: the X-ray Telescope \citep[\xrt;][]{xrt} and the EUV Imaging Spectrometer \citep[\eis;][]{eis}.
\xrt\ provides broadband soft X-ray (SXR) imaging of hot thermal plasma in the corona, while \eis\ provides EUV slit spectroscopy observations of the solar transition region and corona across the spectral ranges 170\,--\,210\,\AA\ and 250\,--\,290\,\AA.

During the March and April windows, \hinode\ ran HOP~476\footnote{\url{https://www.isas.jaxa.jp/home/solar/hinode_op/hop.php?hop=0476}}.
This included two observing modes for \xrt.
The first took pairs of images with the thin and medium beryllium filters (Be-thin, Be-med) at 1-minute cadence.
This was designed to provide context on the active region outside flaring intervals, and was run by default.
The other was optimised for flare observations and was initiated by \xrt's flare trigger.
This mode took ten Be-thick (thick beryllium filter) images followed by a Be-med image.
While the Be-thick filter would capture the hottest flare plasma ($\sim$5\,MK) on the fastest timescales, the Be-med images would provide lower cadence context on the dimmer, cooler material.
It would also provide a way of estimating temperature from Be filter ratios.
The programme was set to run as fast as possible.
This led to an irregular cadences, because the exposure time was determined by \xrt's automatic exposure control, which varied with the intensity evolution of the flare.
When this mode was triggered on March 19, the programme achieved cadences in the range 3\,s\,--\,20\,s, with a median cadence of 5\,s.

Meanwhile, \eis\ ran the \textsf{Flare266\_Hunter01} study, which, when activated by \xrt's flare trigger, switches to the \textsf{Flare\_SNS\_v2} sit-and-stare study.\\
\textsf{Flare266\_Hunter01} is a low-telemetry study that uses the 266\as\ slit to take images in the \ion{He}{ii} 256.32~\AA\ line at a 57~s cadence.
The response study uses the 2\as\ slit and obtains 17 spectral windows featuring a wide range of lines with formation temperatures from 0.05~MK to 20~MK.
The exposure time was 10~s and the cadence was 16.1~s.
This study was previously used in a study of a flare observed by \hinode\ and \solo\ on 2022 April 2 \citep{janvier:2023}.


\hinode\ ran HOP~489\footnote{\url{https://www.isas.jaxa.jp/home/solar/hinode_op/hop.php?hop=0489}} during the October 15 iteration of the Major Flare campaign.
\xrt\ operated in high-cadence (8\,s) mode throughout the window using the Be-thin filter.
For these observations, the flare detection trigger was disabled in order to guarantee continuous coverage of the target active region at consistent cadence.
Meanwhile, \eis\ ran the \textsf{FLAREDOP\_EIS} study  continuously from 14:05~UT to 23:24~UT.
This is a sit-and-stare study that uses the 2\as\ slit and 10~s exposures, with a cadence of 11.1~s.
The hottest line is \ion{Fe}{xxiii} 263.76~\AA\ and there are additional lines from ions formed in the transition region and corona.

\subsubsection{\iris}
\label{sec:coord-iris}
\iris\ \citep{DePontieu2014,DePontieu2021} provides UV slit spectroscopy observations from the solar photosphere to low corona over the ranges 2785\,--\,2835\,\AA\ (near UV), and 1332\,--\,1358\,\AA\ and 1390\,--\,1406\,\AA\ (far UV).
This enables it to provide high resolution line-of-sight velocity (2.7 km~s$^{-1}$ per pixel) and temperature (logT[K]$\approx$~3.5--7) diagnostics.
Simultaneously, the \iris\ Slit Jaw Imager (SJI) provides high-resolution (0.33--0.4\as) context images in four individual filters, centred around \ion{C}{ii}~1330\AA, \ion{Si}{iiv}~1400\AA, \ion{Mg}{ii}~k~2796\AA\ and \ion{Mg}{ii}~h~wing 2803\AA.
During the \campaign s, the \iris\ team prioritized high-cadence spectral observations (with exposure times $\lessapprox$ 2s), which are crucial to  study the quick response of the lower atmosphere to the flare energy deposition \citep[e.g.][]{Lorincik2025}.
\iris\ operated slightly different observing modes in the two March windows.
On March 19th, \iris's campaign spanned 2020-03-19T19:39\,UT to 2024-03-20T00:04\,UT, and used the observing mode 3443105703.
This was a large sit-and-stare with an exposure time in near and far UV bands of 2\,s and a cadence of 3.2\,s.
Strong lines from the chromosphere and transition region were observed (C~II~1336\,\AA, Si~IV~1403\,\AA, Mg~II~k~2796\,\AA) as well as the quasi-continuum near 2814\,\AA.
To reduce telemetry and improve the signal-to-noise ratio, onboard spectral and spatial summing was performed.
The spectral measurements were accompanied by single-channel SJI observations with the 2796\,\AA\ filter, with $\approx$~3~s cadence.

On March 23rd/24th, \iris's campaign spanned 2020-03-23T22:35\,UT to 2020-03-24T02:41\,UT, and used the observing mode 4204700235.
This was also a sit-and-stare, but with a much shorter exposure time of 0.3\,s and a cadence 1\,s.
The linelist included C~II~1336\,\AA, 1343\,\AA, Fe~XII~1349\,\AA, O~I~1356\,\AA, Si~IV~1403\,\AA, 2814\,\AA, 2826\,\AA, 2832\,\AA\ and Mg~II~k~2796\,\AA.
As on March 19th, onboard 2-pixel summing was used to save telemetry and increase the signal.
The spectral measurements were accompanied by single-channel SJI observations, this time in the 1330\,\AA\ filter, with a 0.3\,s exposure time and a 10\,s cadence. The field-of-view observed part of the flaring active region. 

For the 2024 October 15 iteration of the Major Flare campaign, \iris\ carried out a range of coordinated rapid-cadence observations.
Beginning at 15:00 UT, IRIS conducted a dense 16-step context raster at 5\,s cadence (observing mode 3420607436).
This was followed by a 2.5 hour large sit and stare observation (observing mode 3443605403) at 3s cadence, observing Mg II k, Si IV 1403 and C II 1334/1335.
During the period of \hrieuv\ observations (18:00 -- 22:00 UT), IRIS performed an extremely rapid cadence sit-and-stare with 0.3s exposures and 1s cadence (observing mode 4204700237), observing the strongest spectral lines, e.g.\ Mg II k, Si IV 1403\,\AA~ and C II 1334/1335\,\AA, as well as the continuum near 2814\,\AA.
To increase the signal-to-noise ratio, the data were spatially summed by a factor of 2.
The sit-and-stare was accompanied by SJI images in the 1330\AA~channel with 8s cadence.
Following this, IRIS performed another 90 minute 3s large sit-and-stare observation (observing mode 3443605403) and finished with a final 16 step sparse context raster (observing mode 3400607439), again at 5\,s cadence.



\subsubsection{Other Earth-orbiting and Ground-based Telescopes}
Some observatories provide full-disk, high-duty-cycle observations with limited customisation of their observing mode.
They therefore required no specific coordination.
Those of particular interest to this campaign include, but may not be limited to, the Expanded Owens' Valley Solar Array \citep[\eovsa;][]{eovsa}, \sdo, and the X-ray Sensor onboard the Geostationary Operational Environmental Satellites (\xrs).
Successful coordination was achieved with these observatories during the March and October campaign windows, when the target active region was visible from Earth.
Additionally, \stereo-A was located 8.9\,--\,10\degree\ ahead of Earth during the March/April windows, and approximately 26.8\degree\ ahead during the October window.

\begin{figure}
    \centering
    \includegraphics[height=1.04\textheight]{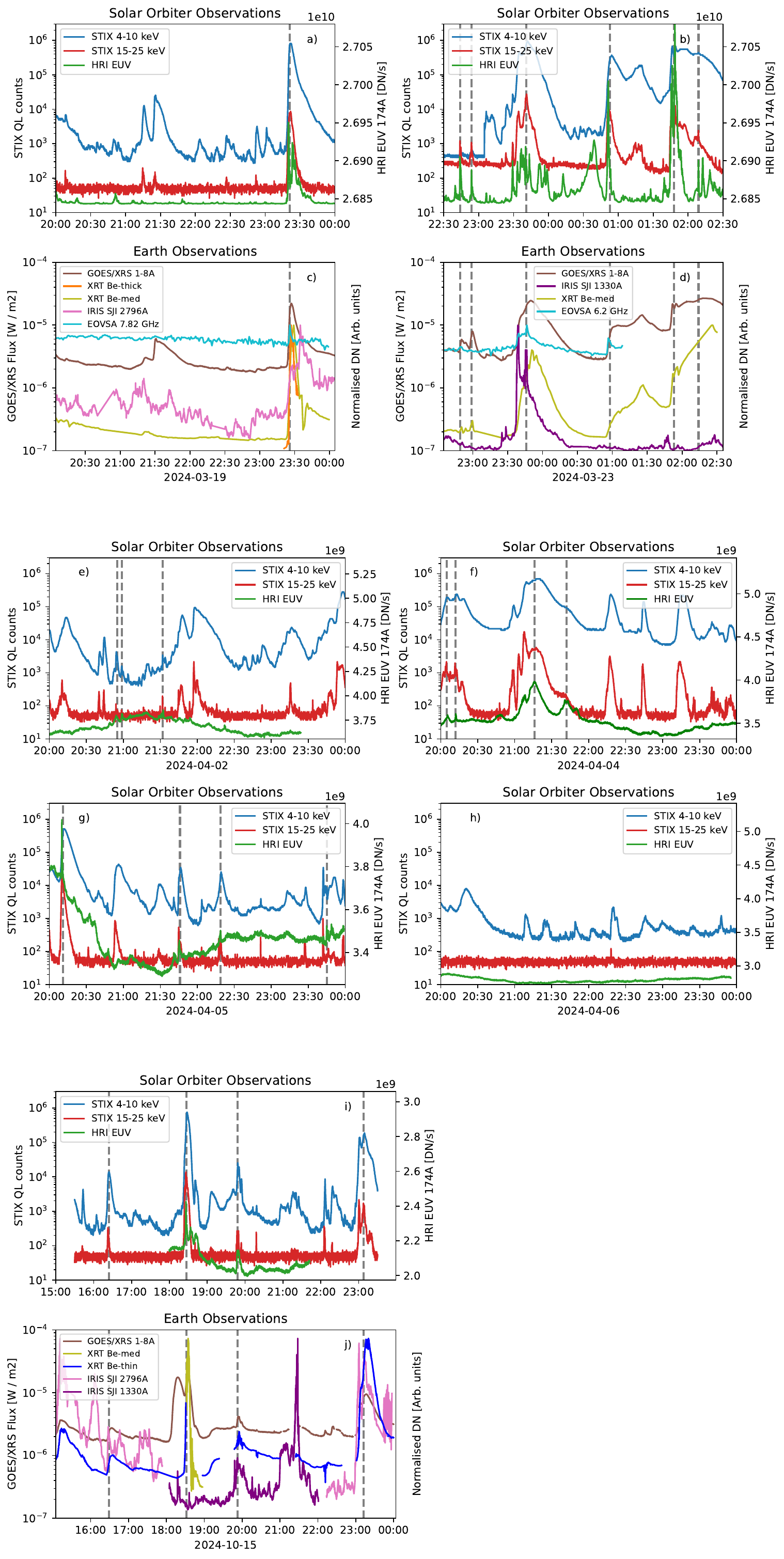}
    \caption{Integrated lightcurves from each observing window of the 2024 \campaign s. Panels a), b), and e)\,--\,i) show \stix\ 4\,--\,10\,keV (blue) and 15\,--\,25\,keV (red), and \hrieuv\ 174\,\AA\ (green) onboard Solar Orbiter.  Panels c), d) and j) show observations from Earth-orbit on March 19\textsuperscript{th},  23\textsuperscript{rd}/24\textsuperscript{th}, and October 15\textsuperscript{th}, respectively.  These include \xrs\ 1\,--\,8\,\AA\ (brown), \hinode/\xrt\ Be-thick (orange), Be-med (yellow) and Be-thin (deep blue), \eovsa\ (cyan), and the \iris\ slit-jaw imager (pink: 2796\,\AA; purple: 1330\,\AA). The time axis of these panels has been shifted to account for the difference in light travel times. Flares observed in the target active region are denoted by the grey vertical dashed lines.}
    \label{fig:lightcurves}
\end{figure}

\begin{table}[]
    \centering
    \caption{Flares jointly observed by \stix\ and \hrieuv\ during the \campaign. Events listed in {\bf bold font} are those have confirmed \spice\ observations of the flare sources. Additional instruments that observed each flare are listed in the right-most column.}
    \begin{tabular}{c|c|c|c}
        Date       &Time\textsuperscript{a} [UT]   &\goes\ class    &Joint observations  \\
        \hline
        2024-03-19 &23:21  &M2.2   &\eovsa, \goes, \iris\ slit, \\
                   &       &       &\sdo, \xrt\ (Be-med \& Be-thick), \eis \\
        \midrule
        2024-03-23 &22:44  &C5.9   &\goes, \iris\ slit, \sdo, \xrt\ (Be-med) \\
                   &22:54  &C8.0   &\goes, \iris\ slit, \sdo, \xrt\ (Be-med) \\
                   &\textbf{23:41}  &\textbf{M2.5}   &\eovsa, \goes, \iris\ slit, \sdo, \\
                   &       &       &\xrt\ (Be-med), \eis\ \\
        \midrule
        2024-03-24 &00:53  &C9.9   &\eovsa, \goes, \iris\ slit, \sdo, \\
                   &       &       &\xrt\ (Be-med) \\
                   &\textbf{01:48}  &\textbf{M2.2}   &\goes, \iris\ slit, \sdo, \xrt\ (Be-med) \\
                   &02:09  &M2.7\textsuperscript{b}   &\goes, \iris\ slit, \sdo, \xrt\ (Be-med) \\
        \midrule
        2024-04-02 &20:55  &$\sim$B5\textsuperscript{c}   &  Not seen from Earth\\
                   &20:59  &$\sim$B3\textsuperscript{c}   & \\
                   &21:32  &$\sim$B6\textsuperscript{c}   & \\
        \midrule
        2024-04-04 &20:05  &C2.5\,$\pm$\,0.3\textsuperscript{c}   & Not seen from Earth\\
                   &20:12  &C5.0\,$\pm$\,0.6\textsuperscript{c}   & \\
                   &21:16  &M1.0\,$\pm$\,0.1\textsuperscript{c}   & \\
                   &21:42  &C3.2\,$\pm$\,0.3\textsuperscript{b,c}   & \\
        \midrule
        2024-04-05 &20:11  &C9.9\,$\pm$\,1.1\textsuperscript{c}   & Not seen from Earth\\
                   &21:46  &C1.4\,$\pm$\,0.2\textsuperscript{c}   & \\
                   &22:19  &C1.1\,$\pm$\,0.1\textsuperscript{c}   & \\
                   &23:45  &$\sim$B5\textsuperscript{c}   & \\
        \midrule
        2024-10-15 &\textbf{16:24}\textsuperscript{d} &\textbf{C2.6} & \goes, \iris\ slit, \sdo, \xrt\ (Be-thin), \eis \\
                   &18:27 &M2.1 & \goes, \sdo, \xrt\ (Be-thin), 
                   \eis \\
                   &19:48 &C4.0 & \goes, \sdo, \iris\ slit, \xrt\ (Be-thin) \\
                   &\textbf{23:08}\textsuperscript{d} &\textbf{C9.4} & \goes, \sdo, \iris\ slit, \xrt\ (Be-thin), \eis \\
        
    \end{tabular}
    \label{tab:flares}
    \begin{FlushLeft}
        \footnotesize{\textsuperscript{a}Flare times are approximately given by the time of the main peak as seen by \hrieuv\ and represent the observation time at \solo.\\ \textsuperscript{b}\goes\ class inflated by decay of previous flare.\\ \textsuperscript{c}\goes\ class estimated from \stix\ background detector as flare was not visible from Earth \citep[][]{stiefel:2025}.\\
        \textsuperscript{d}Not observed by \hrieuv.}
    \end{FlushLeft}
\end{table}

\section{Summary of Observations}
\label{sec:obs-summary}
Figure~\ref{fig:lightcurves} shows a summary of lightcurves for observations from each observing window.
Panels a), b) and e)\,--\,h) show observations from \solo, namely \stix\ (4\,--\,10\,keV, blue; 15\,--\,25\,keV, red) and \hrieuv~174\,\AA\ (green).
Panels c), d) and j) represent supporting ground-based and Earth-orbiting observations.
Times represent light arrival time at the spacecraft/telescope.
To aid visual comparison of features, the time axes of panels showing Earth-based observations have been shifted left to account for the difference in light travel time to \solo\ and Earth.
(See the right column of Table~\ref{tab:windows} for median light travel time differences for each observing window.)
The vertical dashed grey lines represent flares observed by \solo\ from the target active regions, 22 in all.
Their peak times, \goes\ classes, and observational coverage are listed in Table~\ref{tab:flares}.
Of these flares, the \spice\ slit captured at least four events (indicated in bold font in Table~\ref{tab:flares}), as well as crossing a microflare within the flaring active region on 23rd March at $\sim$23:40UT.
All 7 flares in the two March windows were captured by the \iris\ slit, \xrt's Be-med filter (1-minute cadence), \aia, \hmi, and \xrs\ (panels c and d, Figure~\ref{fig:lightcurves}).
Additionally, the March 19 M2.2 flare triggered the high cadence \xrt\ Be-thick flare program (orange line, Figure~\ref{fig:lightcurves}c) and \eovsa, which also observed the March 23/24 events during its local observing day, i.e.\ prior to 2024-03-24~01:08\,UT.
The \eis\ slit observed the March 19 M-class flare with the \textsf{Flare\_SNS\_v2} study beginning at 23:24:34~UT.
The intensity of the hottest EIS ion, \ion{Fe}{xxiv}, rises rapidly within 60~s at three distinct locations along the slit, and the emission remains present for around 10~min.
Because \xrt\ failed to trigger on the 23 March flares, the \textsf{Flare\_SNS\_v2} study did not run.
However, the flare was captured in the \textsf{Flare266\_Hunter01} \ion{He}{ii} images.
During the October campaign, \eis\ observed \ion{Fe}{xxiii} emission during the flares at 16:32~UT, 18:33~UT and 23:16~UT.
The 23:16~UT flare shows the most intense emission.

\begin{figure}
    \centering
    \includegraphics[width=0.95\textwidth]{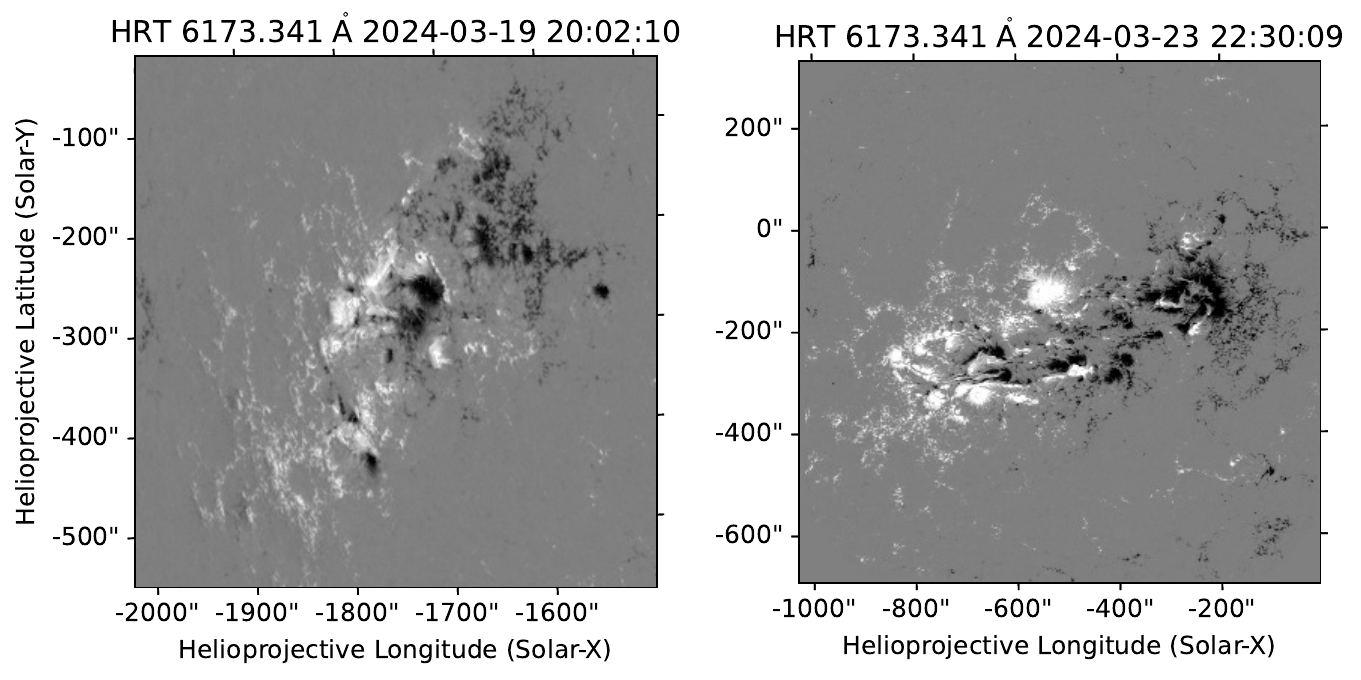}
    \caption{\hrt\ magnetograms of NOAA AR~13615 taken during the March~19\textsuperscript{th} (left) and March~23\textsuperscript{rd} (right) observing windows of the \campaign.}
    \label{fig:magnetograms}
\end{figure}

Magnetograms were successfully produced from \hrt's observations during the March observing windows.
A magnetogram from March~19\textsuperscript{th} (left panel, Figure~\ref{fig:magnetograms}) shows a large, magnetically complex active region, with areas of intermingled strong positive and negative magnetic polarity.
By March~23\textsuperscript{rd} (right panel, Figure~\ref{fig:magnetograms}), the active region has become even more complex due to the emergence of more magnetic flux, and clear magnetic polarity inversion lines are present throughout the active region.
This increased magnetic complexity is coincident with the increased flaring activity observed during the March~23\textsuperscript{rd} observing window.
Magnetograms for the April windows have, so far, proven more challenging due to the proximity of the active region to the limb. 
The increase in the polarimetric noise, foreshortening effects and the variation in the optical aberrations are responsible for a more challenging reduction of the April windows observations and interpretation of the magnetic field vector. 
That said, magnetograms from April 5th have already been successfully used by one study \citep{Tan:2025}, and \hrt's optical continuum images are expected to prove useful context in analysing flares from these windows.

Movies of the full observing windows are provided in the supplementary material.  These include imaging observations from \hrieuv, \iris\ SJI, and the \xrt\ Be-med and Be-thick filters.
Appendix~\ref{app:movies} provides a brief description of these.
While, as previously stated, detailed analysis of the observations will be addressed in future papers, preliminary analysis underscores the scientific value of campaigns such as this one.
In the following subsections, we shall highlight some initial findings.

\begin{figure}
    \centering
    \includegraphics[height=0.75\textheight]{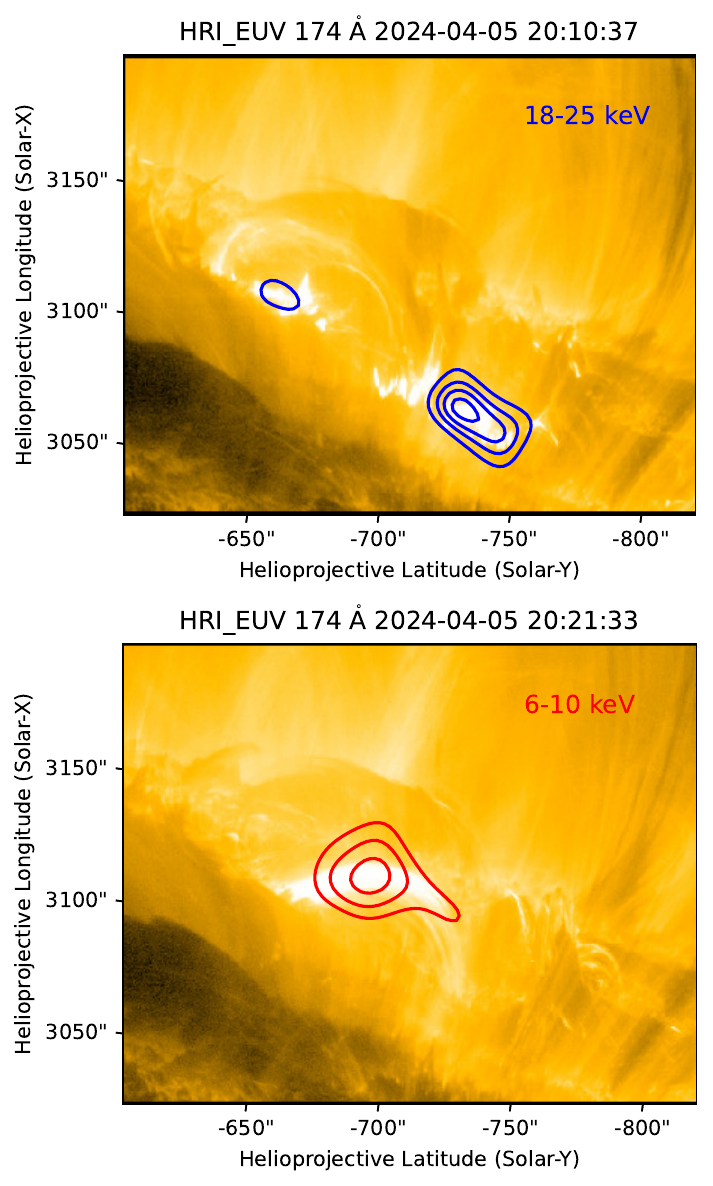}
    \caption{Two \hrieuvcor\ images, taken from the accompanying movie, of the estimated C9.9 flare of April 5\textsuperscript{th}. At 20:10 (top image), the brightest emission is from low altitude footpoint/ribbons sources, while at 20:21 (bottom image), it is from a newly formed flare loop linking the previously bright footpoint regions in the top image.}
    \label{fig:hri_footpoint_and_loop}
\end{figure}

\section{Initial Findings}
\label{sec:results}
\subsection{EUV Emission Associated with Non-thermal and Thermal HXR Sources}
\label{sec:euv-nt-th}
The \hrieuv~174\,\AA\ lightcurves seen in Figure~\ref{fig:lightcurves} are very impulsive.
They tend to more closely match the \stix\ 15\,--\,25\,keV time profiles, which is often associated with non-thermal footpoint emission.
This is confirmed by the accompanying movie of the estimated C9.9 of April 5\textsuperscript{th}.
Figure~\ref{fig:hri_footpoint_and_loop} shows two key frames from the movie taken at 20:10\,UT and 20:21\,UT.
The flare's proximity to the limb makes it easier for us to discern the relative altitudes of different sources.
At 20:10\,UT (top), the brightest emission comes from low altitudes.
At 20:21\,UT (bottom), the low altitude emission has faded, and a bright flaring loop connects the two previously bright footpoint regions.
These observations are consistent with the standard flare model \citep[][]{Carmichael:1964,Sturrock:1966,Kopp:1976}.
This lends the interpretation that the loops were created shortly before the 20:10\,UT via magnetic reconnection, which caused energy to flow down the newly formed magnetic field lines to the chromosphere, presumably via accelerated charged particles.
The denser material at the lower altitudes absorbed this energy, causing it to heat rapidly, giving rise to the intense 174\,\AA\ emission seen at 20:10\,UT.
Plasma ablated higher into the loops must have been too hot to be observed in the 174\,\AA\ passband ($\sim$1~MK) at this time.
However, as the plasma cooled, the flaring loops became clearly identifiable roughly 10 minutes later.
Analyses like this were hampered in the past by limitations in the cadence and saturation levels of solar EUV imagers.

\subsection{\hrieuv\ Reveals Plasma Dynamics in Flares on Scales Not Previously Imaged}
\label{sec:hri-vs-aia}
Figure~\ref{fig:hri_vs_aia}, and its accompanying movie in the supplementary material, show a comparison of \hrieuvcor\ and \aia~171\,\AA\ observations of the March 19\textsuperscript{th} flare.
The upper middle panel of Figure~\ref{fig:hri_vs_aia} shows normalised time profiles of \hrieuvcor\ (blue) and \aia~171\,\AA\ (orange) emission integrated over the same field of view.
The x-axis of the \aia\ lightcurve (top, orange) has been shifted relative to that of the \hrieuv\ lightcurve (bottom, blue) to account for the difference in light travel time.
Thus, we can directly compare features in both lightcurves.
Both show similar features.
However, there is clear additional temporal structure revealed by \hrieuv's higher cadence.
An examination of the accompanying movie unveils greater detail in \hrieuvcor, previously obscured or ambiguous in \aia~\,171\,\AA.
The images in Figure~\ref{fig:hri_vs_aia} highlight two examples.
The left-hand images show \aia~171\,\AA\ (left upper) and \hrieuvcor\ short (left lower middle) and long (left lower) exposure observations corresponding to the first grey dashed vertical line in the flare's impulsive phase.
The \hrieuv\ movie at this time shows finely structured emission in the lower atmosphere, suggestive of energy flowing from the corona shortly after the flare begins.
\aia~171\,\AA\ shows a similar inverted U-shape, but the image is heavily affected by the coarser resolution and severe saturation and blooming, which obscures the dynamics on the finer spatial scales.
This issue is not confined to ribbon emission.
The right panels show similar images corresponding to the second dashed vertical line during the flare's decay phase.
At this time, the emission is dominated by the loops in the flaring arcade, which are clearly visible in the \hrieuvcor\ short exposure image on the right.
Inspection of the accompanying movie shows even more detail, including flows along the arcade.
However, the arcade is completely saturated and bloomed in the \aia~171\,\AA\ image.
Hence, not only is fine scale spatial and temporal structure obscured, but so is whether the observations show emission from energetic ribbons, or decaying thermal loops.
Saturation is also seen in the \hrieuvcor\ normal exposure images.
However, the detectors do not bloom like the AIA images.
This reveals the structure immediately surrounding the saturated pixels, while structure in the saturated region can be determined from the short exposure images.
If desired, the long and short exposure images can be combined to produce non-saturated 21-bit flare images (e.g.\ Movie~12 of supplementary material\footnote{Also available at: \url{https://sidc.be/EUI/data/movie/HRImovies/20240319-HRIEUV-MajorFlareWatch-M2-21bit.mp4}}). 
Flare-optimised \hrieuv\ observations therefore offer a new window for observing and understanding these high energy processes in flares.
A detailed examination of this flare, including these \hrieuv\ observations, is the subject of a separate ongoing study \citep{Hayes:2025_inprep}.

\begin{figure}
    \centering
    \includegraphics[width=0.9\textwidth]{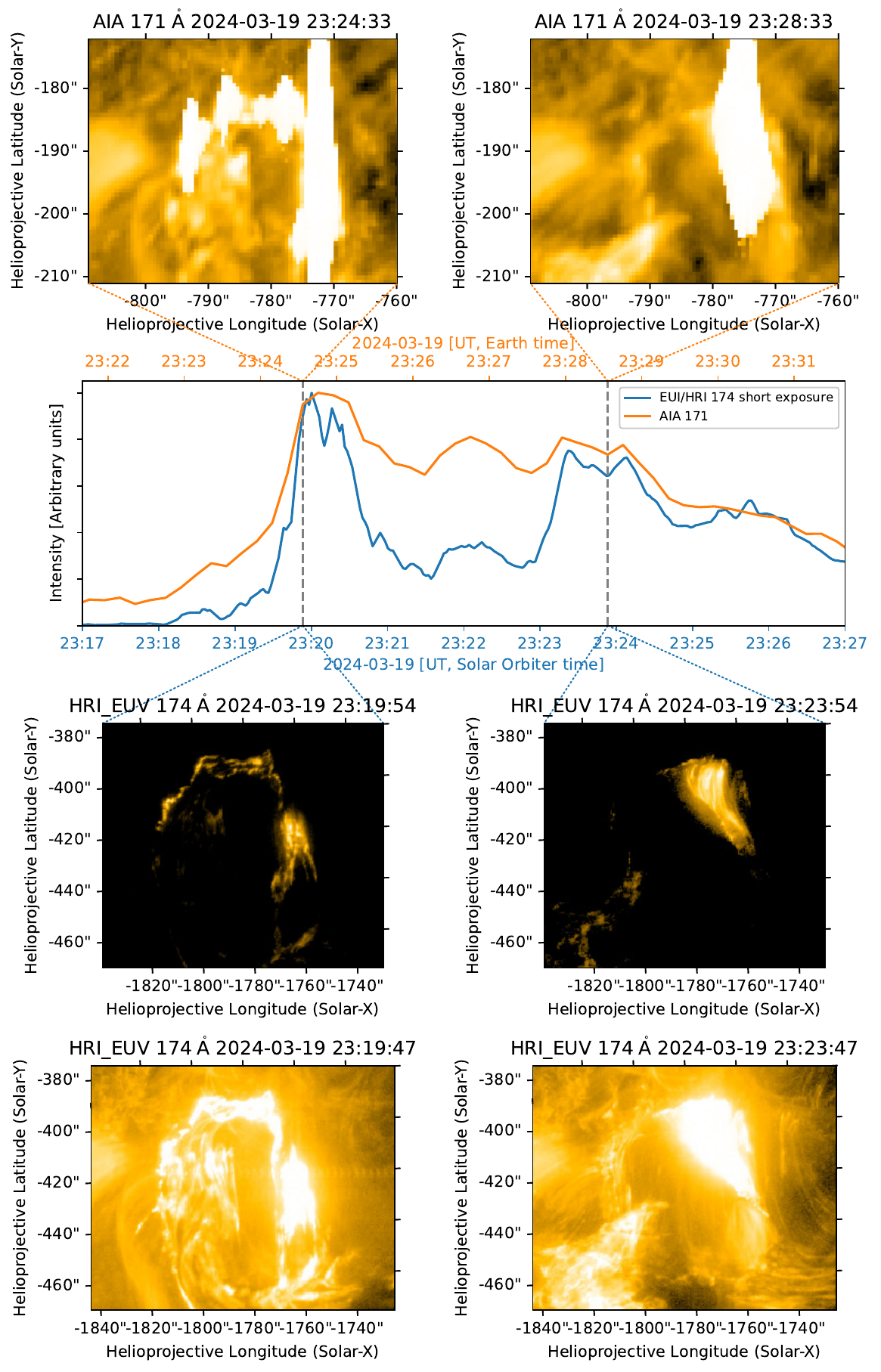}
    \caption{A comparison of \aia~171\,\AA\ and \hrieuvcor\ short exposure flare observations for the March~19\textsuperscript{th} flare. The lightcurves show normalised \hrieuv\ (blue) and \aia\ (orange) emission integrated over the same field of view. The \aia\ time axis (upper, orange) has been shifted relative to the \hrieuv\ time axis (lower, blue) to account for the difference in light travel time. The left and right images correspond to the times of the first and second vertical grey dashed lines on the lightcurves, respectively. The top images show \aia\ observations, while the middle and lower images show short and long exposure \hrieuvcor\ observations, respectively. \aia's lower cadence and saturation/bleeding obscures temporal and spatial evolution during the flare's impulsive phase(s).}
    \label{fig:hri_vs_aia}
\end{figure}

\subsection{Dynamic EUV-Optically Thick Material Complicates Analysis of Limb-events}
\label{sec:euv-optically-thick}
It is important to note that some of the footpoint emission in the top panel of Figure~\ref{fig:hri_footpoint_and_loop} is obscured by optically thick material in front of it.
A prime example of this can be seen in the supplementary movie showing long-exposure \hrieuv\ observations from April 6\textsuperscript{th}.
Although no flare occurred during this observing window, complex dynamics of optically thick material were observed.
The target active region was on the limb, and the bright core of the active region at low altitudes was largely obscured by optically thick material in front of it, which moved up and down in different columns, presumably associated with different magnetic loops.
This included examples of what look like quiescent chromospheric evaporation.
This movement of the optically thick material may raise complications if using \hrieuvcor\ intensity as a proxy for low-altitude emission.
For example, it could introduce apparent temporal variations or a lower total intensity which are not associated with the low-altitude emission, but rather the motions of the optically thick plasma in the foreground.
This finding is consistent with previous observations of EUV low-altitude limb emission made with \trace\ and \aia\ \citep[e.g.][]{ Alissandrakis:2019_err,Alissandrakis:2019}.

\begin{figure}
    \centering
    \includegraphics[width=0.9\textwidth]{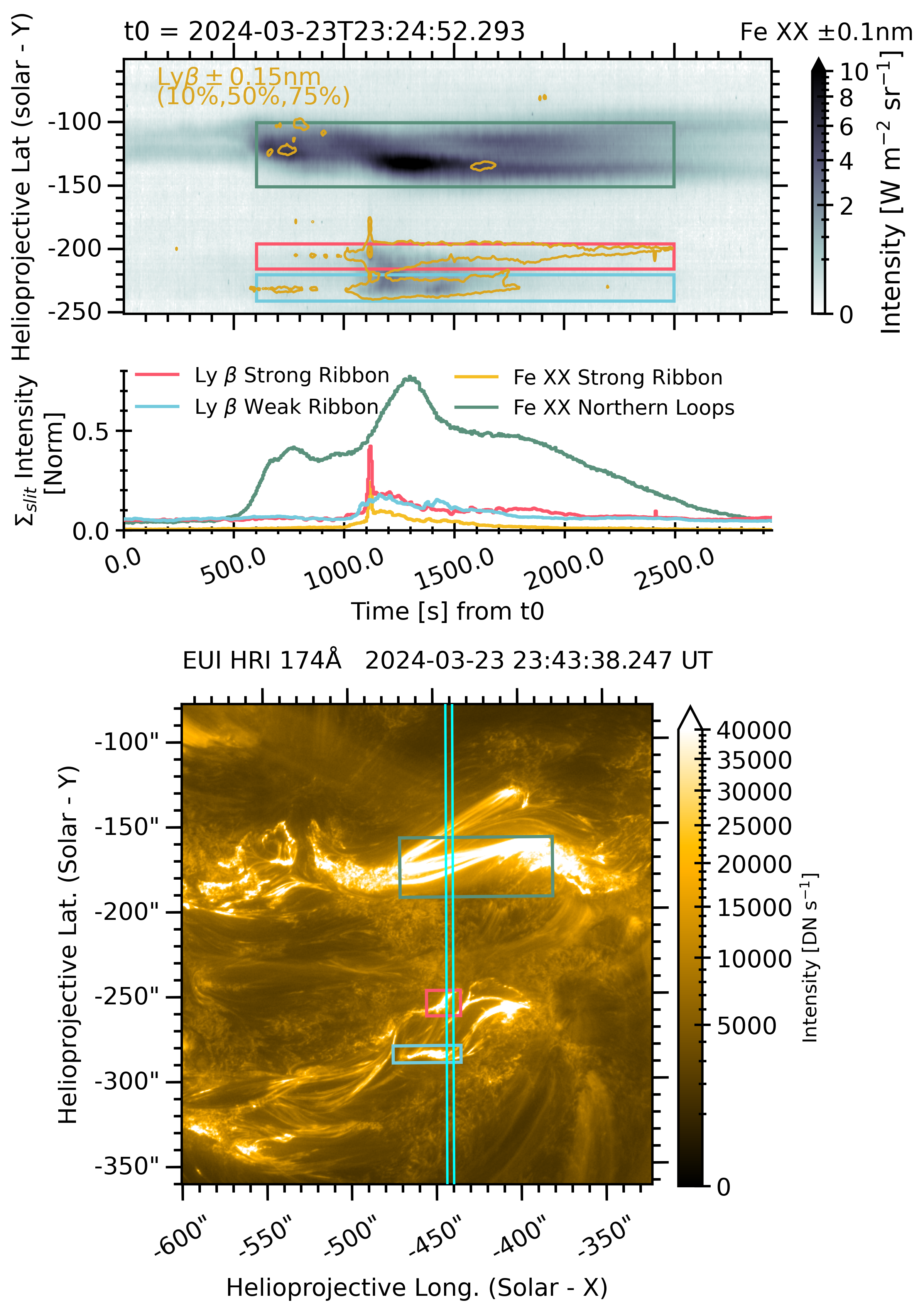}
    \caption{Hot and cool flare emission from the 23rd March 2024, as observed by \spice. In the top panel a space-time map of  Fe~\textsc{xx} ($\sim$10~MK) emission, is shown, with contours representing  H~\textsc{i} Ly~$
    \beta$ ($\sim10$~kK) emission. Both spectral lines were integrated in wavelength within the ranges stated on the figure. Note that the colourbar has been saturated to help highlight less intense features. The SPICE slit cut through flare loops from an M class flare (green box), as well flare ribbons from a smaller flare that occurred nearby a few minutes later (the red box outlines the stronger ribbon, and the blue box the weaker ribbons). Emission summed from the sources indicated by the boxes was normalised to the total emission summed along the field of view shown in the top panel. The stronger flare ribbon is clearly more impulsive in comparison to the weaker flare ribbon. The bottom panel shows a context image from \hrieuv, with northern M flare loops, strong ribbon, and weaker ribbons outlined by boxes of the same colour as in the \spice\ panels. The \spice\ slit is indicated in cyan, which has been shifted $x\sim - 18^{\prime\prime}$ to align with sources on EUI map. No shift has been applied in solar-Y, but an approximate offset of $y\sim50^{\prime\prime}$ between EUI and SPICE is present.  Times are at Solar Orbiter, which was at $\sim0.38$~au.}
    \label{fig:spice_example_image}
\end{figure}

\subsection{EUV Flare Spectroscopy}
\label{sec:spice-results}
At least four flares (2024 March 23rd, $\sim$23:30~UT, 2024 March 24th $\sim$01:40~UT, 2024 October 15 $\sim$16:25~UT and $\sim$23:10~UT) were caught by \spice's slit.
Compact flare ribbons and loops from a microflare were also observed within the flaring active region on 23rd March 2024 $\sim$23:40~UT.
Other events during the campaign show brightenings in the \spice\ lightcurves that may be associated with flares, but detailed analysis is required to confirm their origin. 
Preliminary analysis has so far focused on the observations from March 23 and 24th, and has thus far not revealed signatures of the Orrall-Zirker effect.
However, more detailed analysis of these and/or the event observed on October 15th may yet prove fruitful.
Despite this, these data still provide a wealth of information about the flare plasma. 

The SPICE observations from the 23rd March 2024 event is shown as a space-vs-time image in Figure~\ref{fig:spice_example_image}, which illustrates the powerful combination of simultaneous observations of multi-temperature spectroscopy.
The wavelength-integrated Fe~\textsc{xx} 72.155~nm emission is the background image, and the contours are wavelength-integrated H~\textsc{i} Ly~$\beta$ emission.
In the north of the field-of-view hot (10\,MK) flare loops from an M-class event are observed, which connect two flare ribbons observed by \hrieuv.
A few 10s of arcseconds south, \spice 's slit intersected two compact flare ribbons from a microflare, that brightened a few minutes after the onset of the main M-class flare.
Bright chromospheric emission is obvious, but we also find 10\,MK footpoint emission from Fe~\textsc{xx}. Loops later appear between the two flare footpoints.
Lightcurves from the M flare loops, and the microflare footpoints are shown in the middle panel, illustrating the rapid response of the footpoint emission, and the importance of high-cadence data.
An EUI image is shown for context, with the location of the SPICE slit overlaid.
A rough offset of $\sim18^{\prime\prime}$ has been applied to solar-X based on a comparison of spacetime plots produced with \spice\ and \hrieuv.
However, note that there is an additional offset in solar-Y that has not been corrected in this image. Note that alignment between \hrieuv\ and \spice\ can be achieved with greater fidelity using the SPICE context rasters. 

These observations represent the first time the Ly$\beta$ and Ly$\gamma$ lines have been observed in flares at high-cadence, with spatial information \citep[SDO/EVE previously observed disk-integrated Lyman line emission at lower cadence, e.g.][]{Brown:2016}.
A detailed analysis of these data will feature in a forthcoming study \citep{Kerr:2025_inprep}, which exploits the Lyman decrement (the ratio of Ly$\beta$/Ly$\gamma$) to probe thermal and non-thermal signatures within flare ribbons on small spatial scales.

\section{Discussion and Campaign Insights}
\label{sec:disc}
The \campaign s of 2024 were highly successful.
All instruments operated as expected.
22 flares were observed, including 8 M-class and 3 C9-class.
This represents one of the most comprehensive data sets so far for contributing to one of \solo's top-level science goals: {\it ``How do solar eruptions produce energetic particle radiation that fills the heliosphere?''}.

Preliminary analysis of the campaigns' observations show that many are well-suited for addressing the campaign's first science goal of probing the spatial evolution of flare plasma on timescales relevant to impulsive energy release.
Multiple studies are ongoing on this theme and publications are expected in the future.
Preliminary analysis of the March \spice\ observations has so far failed to find clear signatures of the Orrall-Zirker effect.
Modelling suggests that higher cadence is likely required to capture the transient signal \citep{Kerr:2023}, but more detailed analysis of these and other observations made during these campaigns may yet reveal more promising results.
The campaign's third science goal of obtaining high-energy stereoscopic observations was not met because a flare was not observed from both Earth and \solo\ with a sufficient observer separation angle.
Encouragingly, however, \solo's instruments operated as desired.
Therefore, achieving this science goal is simply a matter of repeating coordinated Earth-\solo\ observing campaigns with the right spacecraft geometry until suitable flares are observed.
Nonetheless, obtaining so many first-of-their-kind, high-quality observations relevant to the first science goal represents an important success, especially considering the long lead time of scheduling decisions, the unpredictability of flares, and the fact that this combination of \solo\ instrument observing modes had never been run together before.
The observations have spawned numerous ongoing scientific studies, some of which have already led to journal submissions \citep[e.g.][]{Tan:2025,Young:2025}.
Additional, less mature studies are expected to lead to publications in the future.
These not only examine topics related to the campaign science goals, but also to additional science questions not originally envisioned.
This is a reflection of the quality and diversity of the campaigns' observations, and demonstrates a force multiplier of the science value of comprehensive flare campaigns.

The campaign was also a success from an operational perspective, and provided lessons for future campaigns.
Collaboration between the instrument teams, the science operations centre and the SOOP coordinator was positive and supportive.
This is important for encouraging new volunteers to become SOOP coordinators, and hence ensuring \solo\ best serves the solar physics community.

\begin{figure}
    \centering
    \includegraphics[width=0.95\textwidth]{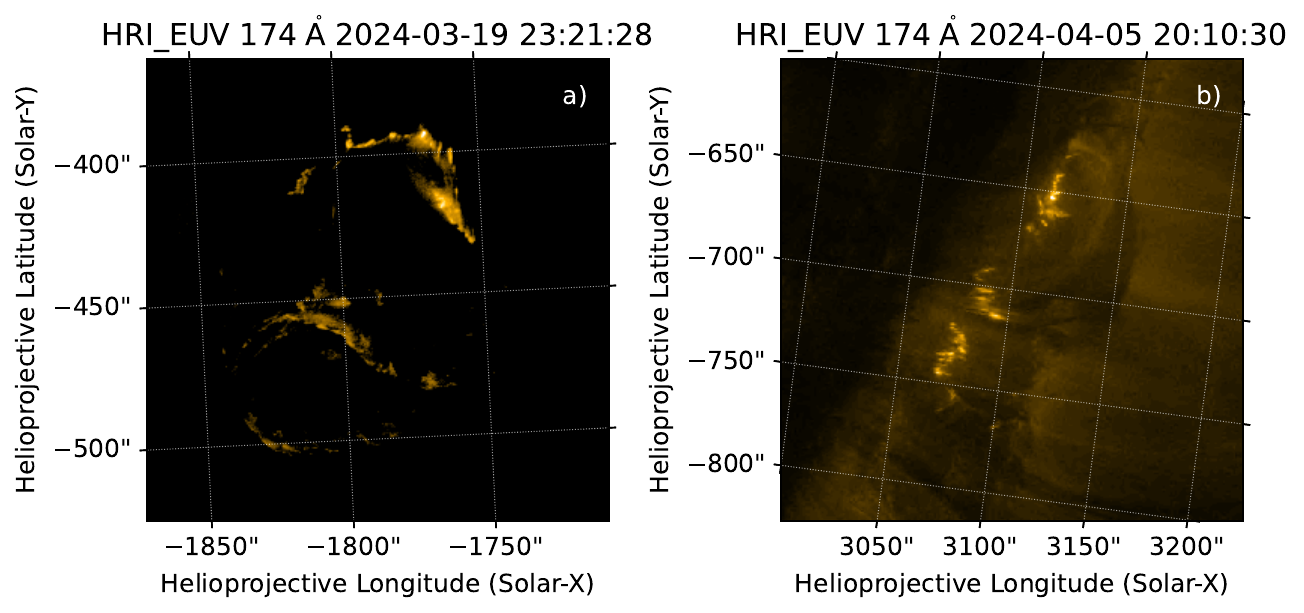}
    \caption{A comparison of \hrieuv\ obtained using the lossless (panel a; 2024 March 19, disk-centre) and lossy (panel b; 2024 April 5, solar limb) compression schemes used during the 2024 \campaign s. Note that low intensity pixels were clipped onboard as part of the lossless scheme, leading to a lot of feature-less regions in panel a). By contrast, no onboard clipping was applied in the lossy scheme, thereby enabling information on dimmer features to be retained in panel b).}
    \label{fig:lossless_vs_lossy}
\end{figure}

Different compression schemes, lossless and lossy, were used for \eui's short-exposure images during the March and April windows, respectively.
Figure~\ref{fig:lossless_vs_lossy} shows a comparison of the two.
The lossless image (Figure~\ref{fig:lossless_vs_lossy}a; March 19, on-disk) provides slightly higher quality images in the bright regions, but discards information of the dimmer regions due to clipping pixels below an intensity threshold.
By contrast, the lossy image (Figure~\ref{fig:lossless_vs_lossy}b; April 5, limb) has greater compression noise, but retains more information on the fainter features.
The lossless scheme was originally justified by the assumption that the brightest features would evolve quickest.
However, preliminary analysis of the lossy images shows that the dimmer features can also evolve on timescales of a few seconds or less.
While the lossless images are visually more attractive, many science cases may be better served by the additional dynamics revealed in the lossy images.
Other examples where the observing modes might be optimised include managing count rates in \spice.
Some of its strong spectral lines (C~\textsc{iii}, O~\textsc{iii}, O~\textsc{vi} 103.2~nm) saturated within flare footpoints, which could be mitigated by reducing the exposure time and/or using the 2$^{\prime\prime}$ slit to reduce the count rates.

Further optimisation could be achieved in coordination with ground-based observatories, which was partially successful.
Simultaneous observations were achieved with \eovsa\ during the March windows.
Coordination with \dkist\ was planned for April due to observing schedule conflicts in March.
However, because the target active region was not visible from Earth in April, this coordination was unfruitful.
From this we conclude that coordination with limited-duty-cycle Earth-based observatories should be prioritised for when \solo\ is near the Sun-Earth line, unless there is a specific science case for multi-perspective observations.
Finally, the choice of observing windows did not allow for coordination with ground-based observatories in Eurasian and African timezones.
Since these provide different capabilities to those in the Americas, a mixture of observing window times may be beneficial to achieve different combinations of coordinated observations, unless the science goal of the campaign is served by a specific ground-based observatory.

The success of the 2024 \campaign s has proven the feasibility of future campaigns, and the lessons learned provide ample scientific justification for future follow-up campaigns.
The next \campaign\ has been scheduled for the remote sensing window of March/April 2025, and has adapted to these lessons.
Moreover, the next campaign will provide another opportunity to obtain stereoscopic flare observations.

The 2024 \campaign s were also a proof-of-concept for a high cadence, high resolution, flare-optimised EUV imager.
The observations achieved with \hrieuv\ during this campaign reveal fine-scale, fast dynamics within flares that are normally obscured by the saturation and low cadence of previous EUV imagers.
This demonstrates the value of further observations, not only from future \solo\ campaigns, but also a flare-dedicated mission with the telemetry resources to observe most of the time, and multiple flare-optimised instruments, such as a direct-focusing HXR spectroscopic imager.
Variants of such a mission concept have been proposed in recent years (e.g.\ FOXSI-SMEX, \citealt{foxsi-smex}; FIERCE, \citealt{fierce}; SPARK, \citealt{spark}).
Moreover, the value of such a combination of instruments was demonstated by the first ever dedicated solar flare sounding rocket campaign.
Two rocket experiments~---~Hi-C-Flare (high resolution, flare-optimised EUV imager) and FOXSI-4 \citep[direct-focusing HXR spectroscopic imager;][]{foxsi4}~---~simultaneously observed the decay phase of an M1-class flare on 2024-04-17, less than two weeks after \solo's April flare observations.
The success of both the \solo\ and sounding rocket flare campaigns reinforces the need for a future solar mission that can make such HXR and EUV spectroscopic and imaging flare observations routinely, and thereby provide the next breakthroughs in our understanding of magnetic reconnection, particle acceleration, and how solar energetic particles populate the heliosphere.
Moreover, such a mission would be further enhanced by EUV imaging spectroscopy that overcomes the challenge that most flares, including in this campaign, are missed by single slit spectrometers.
This will be partially resolved by multi-slit designs like the MUlti-slit Solar Explorer \citep[MUSE; ][]{muse}.
However, maturing technology will soon enable designs that facilitate full 2D EUV imaging spectroscopy \citep[e.g.\ ][]{sisa}, which will exponentially enhance the science return in the study of the EUV signatures of solar flares.

\section{Conclusions} 
\label{sec:conclusion} 
The 2024 \campaign s were highly successful, making observations of 22 flares from B-class to M-class.
These represent the first coordinated \solo\ observations of large flares.
They have provided a rich dataset for future detailed studies of solar flare processes that will help \solo\ address one of its top level science goals: {\it ``How do solar eruptions produce energetic particle radiation that fills the heliosphere?''}
The \hrieuv\ observations represent higher spatial resolution, higher cadence, non-saturated EUV images of coronal flare plasma than previously achieved, and have revealed structure and dynamics on spatial and temporal scales never before seen.
Meanwhile, the \spice\ observations have provided valuable insights for future attempts to identify evidence of accelerated ions via the Orrall-Zirker effect.

This paper has demonstrated the operational viability of future \campaign s, and has outlined some important lessons for their optimisation.
Such campaigns, including the one planned for March and April 2025, will provide different combinations of observing modes and coordinations with different Earth-based observatories.
Moreover, they will provide different combinations of viewing angles that will enable further science questions to be addressed.

Despite the many reasons for optimism, \solo\ will only ever provide limited opportunities for flare campaigns due to its operational constraints.
The \hrieuv\ observations, in particular, have demonstrated the need for routine high spatial resolution, high cadence, non-saturated flare observations.
Future mission concepts should be considered that combine such observations with diagnostics of particle acceleration and plasma heating via higher sensitivity and dynamic range HXR imaging spectroscopy.
Such missions promise to provide the next substantial leaps in our understanding of solar flares and eruptions.

In the meantime, this overview of the 2024 \campaign s provides valuable context for ongoing and future studies of the campaigns' observations.
The observations from this campaign will be openly available to the community on the Solar Orbiter Archive (SOAR)\footnote{https://soar.esac.esa.int/soar/}.

\begin{acks}
Solar Orbiter is a space mission of international collaboration between ESA and NASA, operated by ESA.
We are grateful to the ESA SOC and MOC teams for their support.
The STIX instrument is an international collaboration between Switzerland, Poland, France, Czech Republic, Germany, Austria, Ireland, and Italy.
The EUI instrument was built by CSL, IAS, MPS, MSSL/UCL, PMOD/WRC, ROB, LCF/IO with funding from the Belgian Federal Science Policy Office (BELSPO/PRODEX PEA 4000112292 and 4000134088); the Centre National d’Etudes Spatiales (CNES); the UK Space Agency (UKSA); the Bundesministerium für Wirtschaft und Energie (BMWi) through the Deutsches Zentrum für Luft- und Raumfahrt (DLR); and the Swiss Space Office (SSO).
The development of SPICE has been funded by ESA member states and ESA. It was built and is operated by a multi-national consortium of research institutes supported by their respective funding agencies: STFC RAL (UKSA, hardware lead), IAS (CNES, operations lead), GSFC (NASA), MPS (DLR), PMOD/WRC (Swiss Space Office), SwRI (NASA), UiO (Norwegian Space Agency).
The German contribution to SO/PHI is funded by the BMWi through DLR and by MPG central funds. The Spanish contribution is funded by AEI/MCIN/10.13039/501100011033/ and European Union “NextGenerationEU”/PRTR” (RTI2018-096886-C5, PID2021-125325OB-C5, PCI2022-135009-2, PCI2022-135029-2) and ERDF “A way of making Europe”; “Center of Excellence Severo Ochoa” awards to IAA-CSIC (SEV-2017-0709, CEX2021-001131-S); and a Ramón y Cajal fellowship awarded to DOS. The French contribution is funded by CNES.
Hinode is a Japanese mission developed and launched by ISAS/JAXA, with NAOJ as domestic partner and NASA and STFC (UK) as international partners. It is operated by these agencies in co-operation with ESA and NSC (Norway).
HC is supported by the Swiss National Science Foundation Grant 200021L\_189180 for STIX.
Data analysis for this paper was facilitated by the astropy \citep{astropy:2022}, sunpy \citep{sunpy_joss:2020,sunpy_apj:2020,sunpy:2023}, ndcube \citep{ndcube_apj:2023,ndcube_joss:2023}, sunraster, sospice, and fiasco Python packages, and by the SolarSoftware (SSW) distribution. 
IRIS is a NASA small explorer mission developed and operated by LMSAL with mission operations executed at NASA Ames Research Center and major contributions to downlink communications funded by ESA and the Norwegian Space Centre.
\end{acks}


\appendix   

\section{Accompanying Movies} 
\label{app:movies}

\subsection{Observing Summary Movies}
Movies~1\,--\,9 show imaging observations spanning each of the 2024 \campaign\ observing windows.
Movie~1 shows the 2024 March, 19\textsuperscript{th} window.
The top panels show observation from Earth-orbiting assets including \xrt\ Be-thick (left panel), \xrt\ Be-med (middle), and \iris~2796\,\AA\ (right).
The bottom row show the long (left) and short (right) exposure time observations from \hrieuv\ observations.
For consistency, the timestamps of the near-Earth observations have been shifted to the light-arrival time at \solo.
\xrt\ Be-thick observations were only available between 23:20:58\,UT and 23:32:52\,UT Earth-light-arrival time.
That panel is therefore black outside those times.
The \xrt\ Be-med images show low cadence context on the evolution of hotter plasma in the active region, and many brightenings can be matches with brightenings in the \hrieuv\ movies.
The \iris\ movie shows where its slit was pointing.
During the time of the 19\textsuperscript{th} M-class flare, the ribbons can clearly be seen intersecting the slit.
Meanwhile, much more detail can be seen in the \hrieuv\ movies.
This is especially true during the March 19 M-class flare, where the shorter exposure images in particular show ribbon and loop brightenings, structure and flows on timescales of seconds.
Due to the lossless compression and imaging clipping, all but the brightest areas are black in the short-exposure images.
However, by combining them with the long-exposure images, a fuller picture of the flare and wider active regions is obtained.

Movie~2 shows the 2024 March 23\textsuperscript{rd}/24\textsuperscript{th} observations.
Like Movie~1, the bottom rows show the long (left) and short (right) exposure \hrieuv\ movies.
However, as the \xrt\ flare mode was not triggered during this window, there is no Be-thick panel.
The top left panel therefore shows the \iris~1330\,\AA\ slit jaw movie, while the top left panel shows the \xrt\ Be-med movie.
Also like Movie~1, the timestamps of the near-Earth observations have been shifted to \solo-light-arrival time for consistency with the \eui\ observations.
The increased levels of activity during this window is evident from the continuous emission visible in the short-exposure images, despite the same lossless compression and imaging clipping used during the March 19\textsuperscript{th} window.
Flare ribbons are clearly visible in the \iris\ slit-jaw movie.
For the most part, the slit captured the loops joining the ribbons, rather than the ribbons themselves.
However, at certain times it may have captured the base of a set of loops in the north.

Movies~3 and 6 show the 2024 April 2\textsuperscript{nd} and 6\textsuperscript{th} windows, respectively.
As the active region was not visible from Earth, no near-Earth observations are included.
Furthermore, as no large flares occurred during these window, only the long-exposure images are show.
Despite this, many brightenings and plasma flows are seen across the field of view.

Movies~4 and 5 shows the 20204 April 4\textsuperscript{th} and 5\textsuperscript{th} windows, respectively.
Once again, only \hrieuv\ observations are shown, but this time they include the short exposures.
Here we can see the effect of the different lossy compression scheme used compared to the March windows.
The images are clearly have greater compression noise, but nonetheless contain much more information of the dimmer plasma on much higher timescales.
This is particularly helpful in revealing the dynamics of the fainter plasma in the flares.

Finally, Movie 7 shows long exposure \hrieuv\ observations from the October 15\textsuperscript{th}, revealing what appears to be a series of interchange reconnection events that drive plasma flows into the upper corona and possibly beyond.
Movies 8 and 9 shows supporting \xrt\ observations of the same events made with the Be-thin and Be-med filters.
Note that the Be-thin observations span a longer time range than the others.
Also note that, unlike other movies associated with this paper, the timestamps on these \xrt\ movies represent light arrival time at Earth.

\subsection{Comparing \hrieuv\ and \aia\,171\,\AA}
\label{app:hri-vs-aia}
Movie~10 shows comparison of \hrieuv\,174\,\AA\ and \aia\,171\,\AA\ observations discussed in Section~\ref{sec:hri-vs-aia}.
It shows observations of the 2024 March 19 M2.2 flare.
The top left and top right panels show the long and short exposure \hrieuv\ movies respectively, while the bottom panel shows \aia\,171\,\AA.
Many more details of the flare ribbon, loop structures, and plasma flow are revealed by \hrieuv's higher cadence, spatial resolution, and lack of severe saturation and blooming.

\subsection{Dynamic EUV-Optically Thick Material Obscuring EUV Emission}
\label{app:euv-optically-thick}
Movie~11 shows observations discussed in Section~\ref{sec:euv-optically-thick}, taken during the 2024 April 6\textsuperscript{th} observing window.
Hot plasma emits EUV at the limb.
However, it is often obscured by cooler, optically thick material in front of it.
The optically thick material is highly dynamic, rising and falling in what, in some cases, appears to be partial quiescent chromospheric evaporation.
This behaviour may be worthy of additional study as it may enhance our understanding of plasma dynamics in quiescent active regions.
This movie also emphasises the warning in Section~\ref{sec:euv-optically-thick} that the motions of such optically thick material may introduce apparent temporal behaviour in the EUV emission that could be mistaken as characteristic of the EUV-emitting plasma behind it.

\subsection{Combining \hrieuv\ Long and Short Exposure Images}
\label{app:eui_combined_images}
Movie~12 shows an \hrieuv\ movie of the 2024-03-19 M2 flare whose frames have been formed by combining long and short exposure images.
This enables a far greater dynamic range than could otherwise be achieved with a single exposure time.

\begin{authorcontribution}
DFR was the SOOP coordinator of the 2024 March/April \campaign\ and led the writing of the paper.
LAH and GSK were major contributors to writing the paper.
HC and ARI is/was SOOP coordinators of the 2025 March/April and 2024 October campaigns, respectively, and also contributed to the text of the paper.
DW, APW, HJ, and DM supported coordination of the campaigns as members of ESA's \solo\ team.
EK, DB and CV designed the specific \hrieuv\ operations cycle used in the campaigns.
PRY and TAK provided support on the \spice\ aspects of the campaigns and paper, while SK and MZS provided similar with regard to \stix.
DC processed the \hrt\ observations.
KKR and SS enabled \hinode\ coordination for the campaigns, while VP facilitated \iris\ coordination.
\end{authorcontribution}

\begin{fundinginformation}
 GSK acknowledges support from a NASA Early Career Investigator Program award (grant \# 80NSSC21K0460), and a NASA Heliophysics Supporting Research award (grant \# 80NSSC21K0010). LAH was supported by an ESA research fellowship during the majority of this work. She is now supported by a Royal Society-Research Ireland University Research Fellowship. ARI, PRY, and TAK were supported by SO/SPICE funding at NASA GSFC. For ARI these were provided via cooperative agreement 80NSSC21M0180. VP acknowledges support from NASA under contract NNG09FA40C ({\it IRIS}).
 The development of SPICE has been funded by ESA member states and ESA. It was built and is operated by a multinational consortium of research institutes supported by their respective funding agencies: IAS (Centre National d'Etudes Spatiales (CNES), operations lead), STFC RAL (the UK Space Agency (UKSA), hardware lead), GSFC (NASA), MPS (Deutsches Zentrum für Luft- und Raumfahrt (DLR)), PMOD/WRC (Swiss Space Office (SSO)), SwRI (NASA), UiO (Norwegian Space Agency).
\end{fundinginformation}                  

\begin{dataavailability}
All \solo\ data is/will be available through the Solar Orbiter Archive (SOAR; \url{https://soar.esac.esa.int/soar/})
Additionally, \eui\ is released as part of EUI Data Release 6 \citep[][]{euidatarelease6}.
\end{dataavailability}

\begin{materialsavailability}
\end{materialsavailability}

\begin{codeavailability}
\end{codeavailability}

\begin{ethics}
\begin{conflict}
The authors declare that they have no conflicts of interest.
\end{conflict}
\end{ethics}

\bibliographystyle{spr-mp-sola}
\bibliography{main}  


\end{document}